\documentclass[10pt,twocolumn,english,prd,superscriptaddress,nofootinbib,preprintnumbers,showpacs,floatfix]{revtex4-1}

\usepackage[utf8]{inputenc}
\usepackage{amsmath}
\usepackage{amsfonts}
\usepackage{amssymb}
\usepackage{graphicx} 
\usepackage[caption=false]{subfig}
\graphicspath{ {figures/} }
\usepackage[usenames,dvipsnames]{xcolor}
\usepackage{hyperref}   
\hypersetup{
    colorlinks=true, 
    citecolor=MidnightBlue,
    linkcolor=MidnightBlue,
    urlcolor=Cyan}
\usepackage{tikz}
\usepackage{verbatim} 
\usepackage{soul} 
\definecolor{purple}{rgb}{1,0,1}
\definecolor{lime}{HTML}{A6CE39} 

\newcommand{\orcidicon}{%
	\begin{tikzpicture}
	\draw[lime, fill=lime] (0,0) 
		circle [radius=0.16] 
		node[white] {{\fontfamily{qag}\selectfont \tiny ID}};
	\draw[white, fill=white] (-0.0625,0.095) 
		circle [radius=0.007];
	\end{tikzpicture}	\hspace{-2mm}
}
\newcommand\orcidEdnaldo{{\href{https://orcid.org/0000-0002-9388-8373}{\orcidicon}}}
\newcommand\orcidFrancisco{{\href{https://orcid.org/0000-0002-9388-8373}{\orcidicon}}}
\newcommand\orcidManuel{{\href{https://orcid.org/0000-0001-8586-0285}{\orcidicon}}}
\newcommand\orcidTarciso{{\href{https://orcid.org/0009-0007-0450-2672}{\orcidicon}}}
\newcommand\orcidHenrique{{\href{https://orcid.org/0000-0001-7565-4277}{\orcidicon}}}
\newcommand\orcidLuis{{\href{https://orcid.org/0009-0009-4322-6484}{\orcidicon}}}
\begin{document}

\title{Black hole solutions in Cotton gravity coupled to nonlinear electrodynamics}


        \author{Ednaldo L. B.
        Junior\orcidEdnaldo\!\!} \email{ednaldobarrosjr@gmail.com}
\affiliation{Faculdade de F\'{i}sica, Universidade Federal do Pará, Campus Universitário de Tucuruí, CEP: 68464-000, Tucuruí, Pará, Brazil}

	\author{Jos\'{e} Tarciso S. S. Junior\orcidTarciso\!\!}
 \email{tarcisojunior17@gmail.com}
\affiliation{Faculdade de F\'{\i}sica, Programa de P\'{o}s-Gradua\c{c}\~{a}o em 
F\'isica, Universidade Federal do 
 Par\'{a},  66075-110, Bel\'{e}m, Par\'{a}, Brazil}

	\author{Francisco S. N. Lobo\orcidFrancisco\!\!} \email{fslobo@ciencias.ulisboa.pt}
\affiliation{Instituto de Astrof\'{i}sica e Ci\^{e}ncias do Espa\c{c}o, Faculdade de Ci\^{e}ncias da Universidade de Lisboa, Edifício C8, Campo Grande, P-1749-016 Lisbon, Portugal}
\affiliation{Departamento de F\'{i}sica, Faculdade de Ci\^{e}ncias da Universidade de Lisboa, Edif\'{i}cio C8, Campo Grande, P-1749-016 Lisbon, Portugal}

	\author{Manuel E. Rodrigues\orcidManuel\!\!}
	\email{esialg@gmail.com}
	\affiliation{Faculdade de F\'{\i}sica, Programa de P\'{o}s-Gradua\c{c}\~{a}o em 
F\'isica, Universidade Federal do 
 Par\'{a},  66075-110, Bel\'{e}m, Par\'{a}, Brazil}
\affiliation{Faculdade de Ci\^{e}ncias Exatas e Tecnologia, 
Universidade Federal do Par\'{a}\\
Campus Universit\'{a}rio de Abaetetuba, 68440-000, Abaetetuba, Par\'{a}, 
Brazil}


\author{Luís F. Dias da Silva\orcidLuis\!\!} 
        \email{fc53497@alunos.fc.ul.pt}
\affiliation{Instituto de Astrof\'{i}sica e Ci\^{e}ncias do Espa\c{c}o, Faculdade de Ci\^{e}ncias da Universidade de Lisboa, Edifício C8, Campo Grande, P-1749-016 Lisbon, Portugal}


 \author{Henrique A. Vieira\orcidHenrique\!\!} \email{henriquefisica2017@gmail.com}
\affiliation{Faculdade de F\'{i}sica, Programa de P\'{o}s-Gradua\c{c}\~{a}o em F\'{i}sica, Universidade Federal do Par\'{a}, 66075-110, Bel\'{e}m, Par\'{a}, Brazill}

\date{\LaTeX-ed \today}
\begin{abstract}

The framework of General Relativity (GR) has recently been expanded through the introduction of Cotton Gravity (CG), a theoretical extension proposed by J. Harada. This modified approach integrates the Cotton tensor into the gravitational field equations, naturally encompassing all conventional GR solutions while allowing the cosmological constant to emerge as an integration constant. In this study, we delve into the implications of CG when coupled with nonlinear electrodynamics (NLED), constructing and analyzing three distinct static, spherically symmetric configurations. Our investigation centers on the horizon structure, metric characteristics, and the underlying NLED Lagrangian density of each model. We also confront the theoretical predictions with observational data by comparing the calculated shadow radii of these solutions to constraints imposed by the Event Horizon Telescope’s (EHT) measurements of Sgr A*. The results reveal an extensive spectrum of spacetime geometries, ranging from multi-horizon structures to naked singularities. Furthermore, the agreement between the predicted shadow sizes and EHT observations reinforces the viability of these models in describing the astrophysical image of Sgr A* within certain parameter bounds.
\end{abstract}
\maketitle
\def\HMS{{\scriptscriptstyle{\rm HMS}}}


\section{Introduction}\label{sec1}

For over a century, General Relativity (GR) has been the foundation of gravitational physics, successfully passing experimental tests and providing remarkable predictions \cite{Will:2014kxa}. Its successes include the detection of gravitational waves by LIGO/VIRGO \cite{LIGOScientific:2016sjg,LIGOScientific:2017vwq} and the Event Horizon Telescope’s (EHT) imaging of M87* and Sgr A* \cite{EventHorizonTelescope:2019dse,EventHorizonTelescope:2022wkp}, confirming GR’s predictions about spacetime and black hole shadows. However, despite these achievements, GR faces significant theoretical and observational challenges \cite{Berti:2015itd}.
Theoretically, GR predicts singularities—regions of infinite curvature—inside black holes and at the origin of the universe \cite{Hawking:1970zqf}. It also remains incompatible with quantum mechanics, particularly due to issues of renormalization at high energy scales \cite{Rovelli:2008zza}. Observationally, GR struggles to explain dark matter, a hypothesized non-baryonic component needed to account for the galactic rotation curves \cite{Abdalla:2022yfr}, and dark energy, the exotic source driving the late-time cosmic acceleration, typically modeled via the cosmological constant \cite{Copeland:2006wr}.
To address these challenges, various modifications and extensions of GR have been proposed to advance a more complete theory of gravity \cite{Berti:2015itd}.

A recent addition to the landscape of modified theories of gravity is ``Cotton Gravity'' (CG), proposed by Harada \cite{Harada:2021bte}. Unlike GR, where the Einstein tensor governs gravitational dynamics, CG employs the rank-3 Cotton tensor, naturally incorporating higher-order curvature effects. A key feature of this approach is its invariance under local conformal transformations, analogous to Weyl (Conformal) gravity \cite{Mannheim:1988dj}. Indeed, CG generalizes GR by encompassing all the Einstein field equation solutions, both with and without a cosmological constant, the latter emerging naturally as an integration constant. This characteristic has drawn attention due to its implications for cosmic acceleration, offering an alternative to the conventional dark energy paradigm \cite{Sussman:2023wiw,Harada:2023rqw}.
One of CG's intriguing predictions is a modification to the gravitational potential. When deriving static and spherically symmetric vacuum solutions, Harada identified an additional linear term, which was subsequently proposed as a mechanism to account for the galactic rotation curves without invoking dark matter \cite{Harada:2022edl}. These properties have sparked considerable interest in CG as a potential alternative to GR, driving ongoing research into its viability at cosmological scales \cite{Mantica:2022flg,Sussman:2023wiw,Sussman:2023eep,Mantica:2023ihx,Harada:2023rqw,Mantica:2023ssd,Xia:2024tps,Mo:2024rfq,Feng:2024rnh,Junior:2024lvj,Junior:2024cbb,Gurses:2024ltc,Hazinedar:2024kyj,Gogberashvili:2023wed,Junior:2023ixh,Junior:2024vrv}.

However, CG has not been without controversy. Questions regarding its physical viability have arisen, particularly due to its equivalence to GR’s equations of motion \cite{Bargueno:2021iqg,Harada:2021aid}, the ambiguity of its field equations in specific cosmological models, including Bianchi type-I and FLRW spacetimes \cite{Clement:2023tyx,Sussman:2024iwk,Clement:2024pjl,Sussman:2024qsg}, the lack of a well-defined variational principle \cite{Feng:2024rnh}, and the apparent vanishing of conserved charges for all CG solutions \cite{Altas:2024pil}. While various aspects of these issues have been addressed in ongoing research, the theoretical foundation of CG remains a subject of active debate.
Despite its challenges, CG has yet to be dismissed as an obsolete framework. Instead, it offers an intriguing perspective on cosmic dynamics and may serve as a stepping stone for further theoretical developments \cite{Harada:2021aid,Sussman:2024iwk,Sussman:2024qsg,Xia:2024tps,Mo:2024rfq,Gurses:2024ltc}. Through refinements within its framework or integration into broader theories, CG continues to inspire new directions in the quest for a deeper understanding of gravity.

In a recent study, we investigated black bounce solutions of the Simpson-Visser and Bardeen type within the framework of Cotton Gravity (CG), incorporating nonlinear electrodynamics (NLED) as the matter source \cite{Junior:2024cbb}. By solving the CG field equations, we identified two novel generalizations of these bouncing geometries, analyzing their horizon structure, metric properties, regularity via the Kretschmann scalar, and the characteristics of their corresponding NLED and scalar field sources. NLED has been widely utilized in black hole physics, both as a mechanism for removing singularities \cite{Ayon-Beato:1999kuh,Yajima:2000kw,Bronnikov:2000vy,Dymnikova:2004zc,Balart:2014cga,Culetu:2014lca,Fan:2016hvf}, and in studies of black hole thermodynamics \cite{Rasheed:1997ns,Breton:2004qa, Breton:2007bza, Myung:2007xd,Bokulic:2021dtz,Barrientos:2022bzm}, gravitational lensing and shadows \cite{Uniyal:2023inx,Uniyal:2022vdu,Javed:2022psa}. It has also been explored as a fundamental component in charged wormholes, black bounce geometries, and solitonic spacetime solutions \cite{Bronnikov:2017sgg,Bronnikov:2021uta,Bronnikov:2022bud,Rodrigues:2023vtm, Alencar:2024yvh}.

Building on this foundation, we now extend our exploration of novel gravitational solutions in CG by utilizing the properties of NLED. Our primary objective is to derive static, spherically symmetric configurations governed by CG field equations coupled to an NLED Lagrangian asymptotically recovers Maxwell electrodynamics in the weak field limit. We explore the resulting spacetimes by assessing their horizon structure, metric behavior, and shadow properties. Specifically, we analyze the parameter-dependent shadow radius using an approach akin to \cite{Vagnozzi:2022moj}, comparing theoretical predictions with observational constraints from the Event Horizon Telescope (EHT) to delineate the parameter space where these solutions remain viable within current astrophysical data.

This paper is structured as follows: In Sec. \ref{sec:intro}, we introduce the CG field equations coupled to a purely magnetic matter source described by NLED. Sec. \ref{sec:BHsols} presents the static and spherically symmetric black hole solutions we propose, along with a detailed analysis of these solutions. In Sec. \ref{sec:BHshadows}, we investigate the parameter-dependent shadow radius of the solutions and compare it with the shadow bounds of Sgr A* as estimated by the EHT collaboration. Finally, Sec. \ref{sec:conc} offers a summary of the work and a discussion of the main results. Throughout this study, we adopt a geometrodynamic unit system, setting $G=c=1$.

\section{Cotton Gravity coupled to nonlinear electrodynamics}\label{sec:intro}

We begin our considerations from the CG field equations which, according to Harada \cite{Harada:2021bte}, are expressed as
\begin{equation}
    C_{\alpha\mu\nu}=\kappa^2\nabla_{\beta}T_{\phantom{\mu}\alpha\mu\nu}^{\beta},\label{eq:CGfield}
\end{equation}
with $\kappa^2=16 \pi$, and the 3-rank tensors $C_{\alpha\mu\nu}$ and $T_{\alpha\mu\nu}$ are respectively defined as
\begin{align}
    C_{\alpha\mu\nu} \equiv & \, \nabla_{\mu}R_{\alpha\nu} - \nabla_{\nu}R_{\alpha\mu} \nonumber \\
    & - \frac{1}{6}\left( g_{\nu\alpha}\nabla_{\mu} - g_{\mu\alpha}\nabla_{\nu} \right) {\cal R},\label{eq:CottonTensor}
    \\
    \nabla_{\beta}T_{\phantom{\mu}\alpha\mu\nu}^{\beta} \equiv & \, \frac{1}{2} \left( \nabla_{\mu}T_{\alpha\nu} - \nabla_{\nu}T_{\alpha\mu} \right) \nonumber\\
	&-\frac{1}{6}\left(g_{\nu\alpha}\nabla_{\mu}-g_{\mu\alpha}\nabla_{\nu}\right)T,\label{eq:emTensor}
\end{align}
where $C_{\alpha\mu\nu}$ is the Cotton tensor containing the Ricci tensor $R_{\mu \nu}$ and the Ricci scalar ${\cal R}$, and $T_{\alpha\mu\nu}$ contains the energy-momentum tensor $T_{\mu\nu}$ and its trace $T$. The field equations \eqref{eq:CGfield} are a generalization of GR as they are satisfied by any solution of the Einstein field equations, with or without a cosmological constant; the latter arising as an integration constant. Its exact spherically symmetric neutral vacuum solution\footnote{Note that spacetime configurations
containing a linear term have been obtained in other theoretical contexts (e.g. \cite{Mannheim:1988dj,Klemm:1998kf,Kiselev:2002dx,Grumiller:2010bz,Soroushfar:2015wqa,Ghosh:2015cva})} is 
\begin{equation}
    A(r)=1-\frac{2M}{r}+\gamma r-\frac{\Lambda}{3}r^2,\label{eq:CGvacBH}
\end{equation}
where $M$ represents the local mass, $\Lambda$ is the cosmological constant, and $\gamma$ denotes the CG parameter, describing gravitational effects at large distances. Additionally, the energy-momentum conservation law $\nabla_\mu T^{\mu}_{\phantom{\alpha}\nu} = 0$ naturally occurs when contracting Eq. \eqref{eq:CGfield} with $g^{\mu \nu}$.

Our analysis is restricted to purely magnetic configurations within the context of NLED. In such cases, the functional form of the modified Lagrangian for electrodynamics depends on a single non-vanishing field invariant $F$, expressed as ${\cal L}(F)$ \cite{Novello:1999pg}. Accordingly, the energy-momentum tensor is written as
\begin{equation}\label{eq:EMTensor}
    T_{\mu\nu}=\frac{4}{\kappa^2}(g_{\mu\nu}{\cal L}(F)-{\cal L}_F F_{\mu}^{\phantom{\nu}\alpha}F_{\nu\alpha}),
\end{equation}
where the subscript $_F$ notation is used to represent a derivative with respect to the electromagnetic scalar $F$. The latter is constructed from the electromagnetic field tensor $F_{\mu \nu}=\partial_{\mu}A_{\nu}-\partial_{\nu}A_{\mu}$ via $F=\frac{1}{4}F_{\mu \nu}F^{\mu \nu}$,
where $A_{\mu}$ is the electromagnetic gauge field. By the least action principle, the field equations within the NLED framework are expressed as
\begin{equation}
\nabla_\mu ({\cal L}_F F^{\mu\nu})= \partial_\mu (\sqrt{-g} {\cal L}_F F^{\mu\nu})=0.\label{eq:NLEDfield}
\end{equation}

Given our focus on magnetically charged spacetime geometries, we select a purely magnetic ansatz for the electromagnetic gauge $A_{\mu}=q \cos (\theta) \delta^{\phi}_{\mu}$, where $q$ represents the magnetic charge. Consequently, the electromagnetic field invariant yields the solution
\begin{equation}
F= \frac{q^2}{2 r^4},\label{eq:F}
\end{equation}
which allows the NLED Lagrangian density to be expressed as a radially dependent function. In this regard, we note the usefulness of the relation
\begin{equation}
    {\cal L}_F=\frac{\partial {\cal L}}{\partial r} \bigg(\frac{\partial F}{\partial r}\bigg)^{-1},\label{eq:Consistency}
\end{equation}
through which we assess the consistency of the solutions presented in the following sections.

Let us now consider a generic static and spherically symmetric metric
\begin{equation}\label{eq:metric}
ds^2= A(r) dt^2 - B(r)dr^2 - C(r) d\Omega^2,
\end{equation}
where the generic function $A(r)$ describes the spacetime geometry with an explicit dependence on the radial coordinate $r$, and $d\Omega^2= d\theta^2 + \sin^2 \theta \, d\phi^2$. Our analysis shall be restricted to models that satisfy $B(r)=A^{-1}(r)$ and $C(r)=r^2$. To find spherically symmetric solutions in CG coupled to NLED, we substitute Eq. \eqref{eq:metric} into \eqref{eq:CGfield}, retrieving the following equations of motion:
\begin{eqnarray}
 && \frac{A(r)}{3 r^5}
 \big[r^5 A^{(3)}(r) + r^4 A''(r) - 2r^3 A'(r) + 2r^2 A(r) 
		\nonumber \\  
 && - 4 q^2 r {\cal L}'_{F}(r) + 16 q^2 {\cal L}_{F}(r) + 2 r^5 {\cal L}'(r) - 2r^2\big] = 0,\label{eq:CGfieldeqs1}
\end{eqnarray}
\begin{eqnarray}
 && \frac{1}{6 r^3} \big[ r^5 A^{(3)}(r)+r^4 A''(r)-2 r^3 A'(r)+2 r^2 A(r)
		\nonumber\\   
 && -4 q^2 r  {\cal L}_F'(r)
   +4 q^2 {\cal L}_F(r)-4 r^5 {\cal L}'(r)-2 r^2 \big]= 0,\label{eq:CGfieldeqs2}
\end{eqnarray}
for the $010$ and $212$ components, respectively, and their symmetric for the $001$ and $221$ components. The $313$ component of the field equations is linearly dependent on the $212$ component (i.e., Eq. \eqref{eq:CGfieldeqs2}) via a multiplication factor $\sin (\theta)$, and similarly, the $331$ component is linearly dependent on the $221$ component. Here, we adopt the prime notation $'$ to denote a derivative with respect to the radial coordinate $r$. Solving Eqs. \eqref{eq:CGfieldeqs1} and \eqref{eq:CGfieldeqs2} with respect to ${\cal L}(r)$ and ${\cal L}_F(r)$ yields  
\begin{align}
&{\cal L}(r) = -\frac{r A'(r)+A(r)-1}{2r^2} + \frac{2f_1 q^2}{r} + f_0,\label{eq:Lgeneral}
\\
&{\cal L}_{F}(r) =\frac{r^2 \left(r^2 A''(r) - 2A(r)+2\right)}{4 q^2} + f_1 r^3,\label{eq:LFgeneral}
\end{align}
which verify the consistency condition \eqref{eq:Consistency}, and where $f_0$ and $f_1$ appear as integration constants with units of $M^{-2}$ and $M^{-3}$, respectively. Thus, if $A(r)$ is known, one can obtain the general form of the Lagrangian density ${\cal L}(r)$, which can be expressed as a function of $F$ via Eq. \eqref{eq:F}. Alternatively, solving Equations \eqref{eq:CGfieldeqs1} and \eqref{eq:CGfieldeqs2} with respect to $A(r)$, for a given ${\cal L}(r)$ function, allows new solutions in CG coupled with NLED to be found. This is the strategy we adopt in this paper. First, we propose a NLED Lagrangian density that satisfies the Maxwell electrodynamics in the weak field, drawing from existing NLED models that aim to extend classical electrodynamics to regimes where quantum corrections become relevant (e.g., \cite{Kruglov:2014hpa,Dunne:2004nc}). We supply the proposed Lagrangian to the motion equations \eqref{eq:CGfieldeqs1} and \eqref{eq:CGfieldeqs2}, and find the corresponding metric function $A(r)$. Next, we perform an inverse process in that we supply Eqs. \eqref{eq:CGfieldeqs1} and \eqref{eq:CGfieldeqs2} with the obtained metric function $A(r)$, to determine the general form of the NLED Lagrangian density that generates the aforementioned configuration. Such an approach has frequently been previously employed in the literature \cite{Bronnikov:2017xrt,Bokulic:2023afx}.

\section{BLACK HOLE SOLUTIONS\label{sec:BHsols}}

\subsection{Model I}\label{ssec:I}

The first spacetime geometry we present is obtained by assuming a Lagrangian density in the form ${\cal L}(F) = \frac{F}{1+aF}$, inspired by the NLED models presented in \cite{Kruglov:2014hpa}. Here $a$ is a dimensionless parameter that controls the deviation from linearity. The metric function that satisfies all the components of the CG field equations (i.e., check Eqs. \eqref{eq:CGfieldeqs1}, \eqref{eq:CGfieldeqs2} and the corresponding discussion) is written as
\begin{align}\label{eq:A(r)_I}
   & A(r)= \, 1-\frac{2 M}{r} +\gamma  r - \frac{\Lambda}{3}r^2 + \frac{q}{2} \sqrt[4]{\frac{F}{a}} \times 
\nonumber\\
   & \Bigg[ \arctan \left(S^{-}_a\right) - \arctan \left(S^{+}_a\right) +\frac{1}{2}\log\left(\frac{S^{+}_a\sqrt{F} + 1}{S^{-}_a\sqrt{F} + 1}\right) \Bigg],
\end{align}
where we define $S^{\pm}_a=1 \pm \sqrt[4]{\frac{4}{a F}}$ and resort to Eq. \eqref{eq:F} to further simplify the expression. By inspection of \eqref{eq:A(r)_I}, it is possible to obtain the CG vacuum solution (Eq. \eqref{eq:CGvacBH}) when considering $q \rightarrow 0$. 

To examine the behavior of the metric function \eqref{eq:A(r)_I} we first identify potential horizons through the condition $g^{-1}_{rr} (r_H) = 0$ (which coincides with the Killing horizon $g_{tt}(r_H) = 0$ for every model in this work), where the radius $r_H$ denotes the radial coordinate of an event horizon. Thus, we search for solutions of
\begin{equation}
    A(r_H)=0.\label{eq:hor_cond1}
\end{equation}
Additionally, we assess the nature of such horizons by 
computing the first derivative of $A(r)$
\begin{equation}\label{eq:hor_cond2}
    \left.\frac{dA(r)}{dr}\right|_{r=r_H}=0,
\end{equation}
which signals the presence of a degenerate horizon should it be satisfied as well. Together, Eqs.\;\eqref{eq:hor_cond1} and \eqref{eq:hor_cond2} allow us to determine critical parameter values of a given model which give rise to extreme configurations (i.e., containing at least one degenerate horizon). We note that in this approach, the parameter values are chosen with the aim of exploring the range of possible geometrical structures. This often leads to values well above any constraints found in literature, particularly for the $\Lambda$ and $\gamma$ constants. Throughout all the models in this work, all quantities are normalized with with respect to the mass parameter $M$ (e.g. $q$ is expressed in units of $M$). Such a choice facilitates comparisons between all the considered models, as well as with others such as the Schwarzschild geometry.

\begin{figure}[ht!]
    \includegraphics[width=\columnwidth]{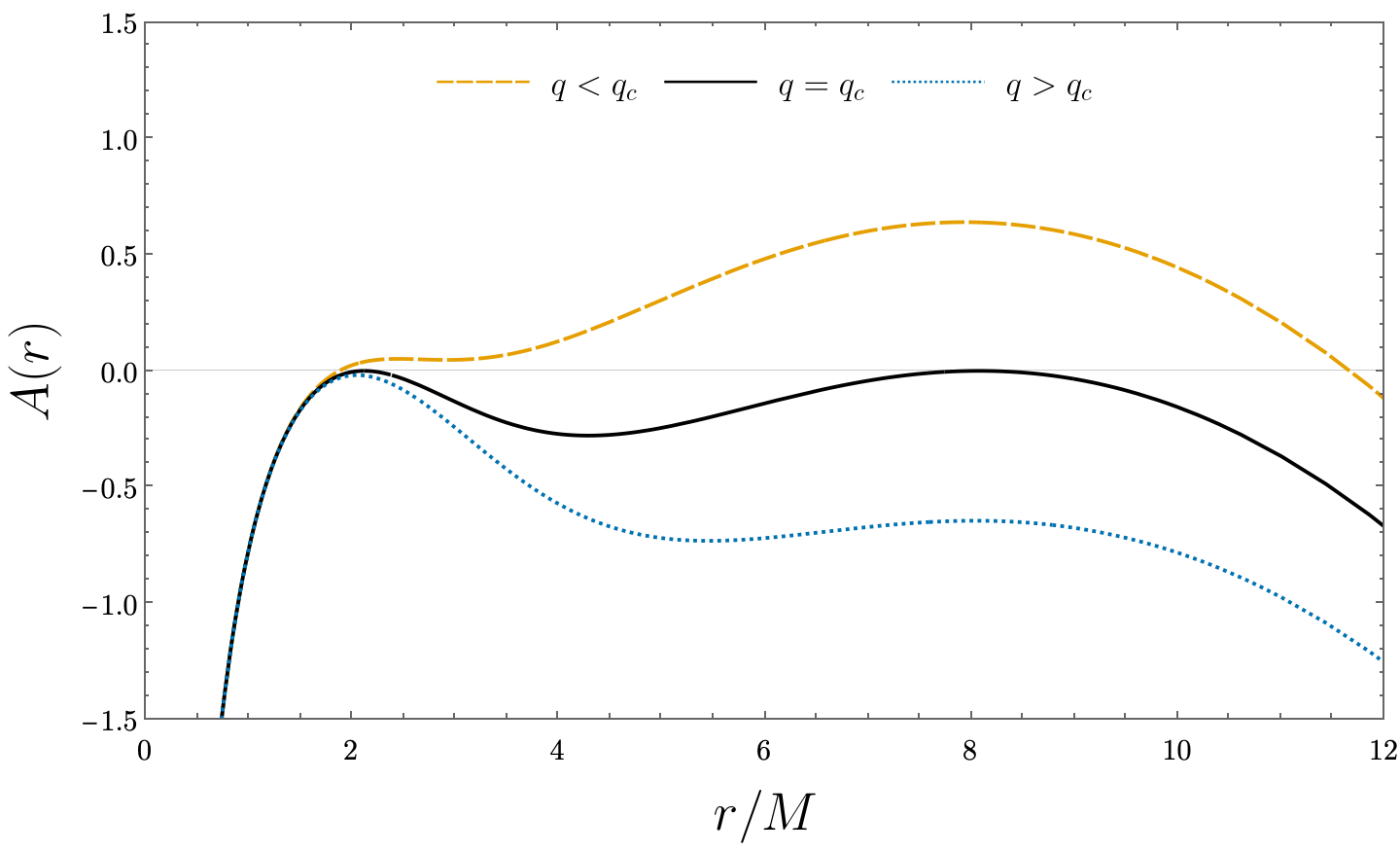}
    \caption{The metric function $A(r)$ of Model I, as described by Eq. \eqref{eq:A(r)_I}, for three distinct charge values: $q < q_c$, $q = q_c$, and $q > q_c$, where $q_c = 9.234 M$. Model parameters: $a=3.0$, $\gamma=0.5 M^{-1}$, $\Lambda = 0.12 M^{-2}$.}
    \label{fig:qcrit_I}
\end{figure}

For Model I, we explore the critical values of $q$ and $a$. In the first configuration, the critical value of the magnetic charge $q_c$ is determined considering the following values for $a=3.0$, $\gamma = 0.5 M^{-1}$ and $\Lambda = 0.12 M^{-2}$. Under such assumptions, the critical charge is found to be $q_c = 9.234 M$. Figure \ref{fig:qcrit_I} illustrates the behavior of the metric function \eqref{eq:A(r)_I} according to the previous assumptions for three distinct charge values: $q > q_c$, $q = q_c$, and $q < q_c$.
When $q = q_c$, the configuration is that of an extreme black hole containing a degenerate double inner horizon at the local maximum close to $r = 2M$, in addition to a degenerate horizon containing the cosmological event horizon near $r=8M$. Below the critical charge value, the geometry contains a single event horizon and a cosmological horizon. When the magnetic charge exceeds the critical value $q > q_c$, the geometry becomes a horizonless configuration of a $(-,+,-,-)$ type, limiting the magnetic charge to $q\leq q_c$ for the considered parameter values. Overall the exact number of horizons present in a given configuration depends on the values of $a, q, \gamma$ and $\Lambda$, yet it is never greater than four (including the cosmological event horizon). Figure \ref{fig:qcrit_I} also shows that the effect of the magnetic charge is more pronounced at $r \gg M$, where the differences between the three curves are more significant. Nonetheless, the asymptotic behavior of this geometry is dominated by $\Lambda$. Note that, should a negatively valued cosmological constant be considered, the cosmological horizon would be absent from all configurations.

\begin{figure}[ht!]
    \includegraphics[width=\columnwidth]{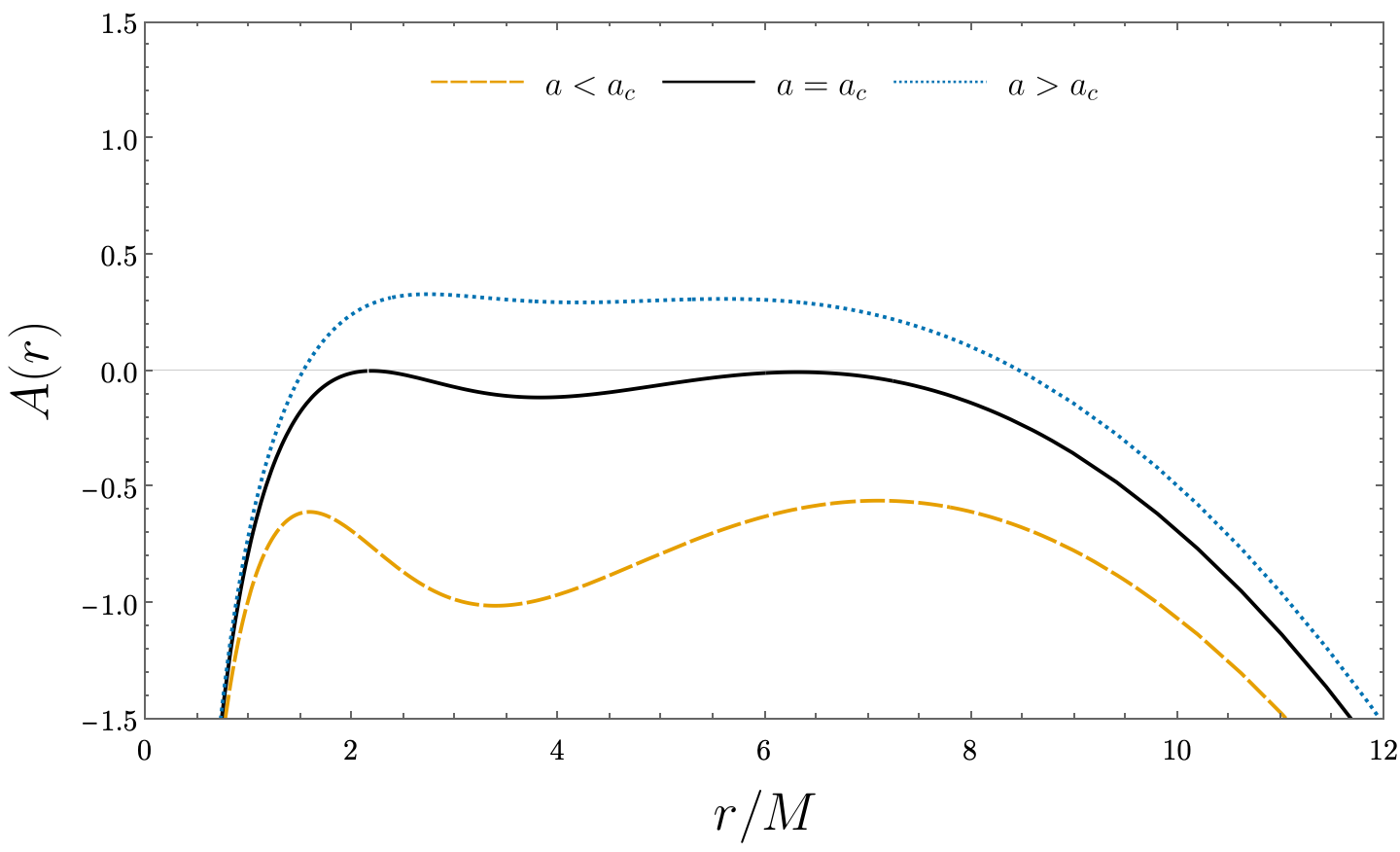}
    \caption{The metric function $A(r)$ of Model I, as described by Eq. \eqref{eq:A(r)_I}, for three distinct charge values: $a < a_c$, $a = a_c$, and $a > a_c$, where $a_c = 3.061$. Model parameters: $q=7.5M$, $\gamma=0.5M^{-1}$, $\Lambda= 0.15 M^{-2}$.}
    \label{fig:acrit_I}
\end{figure}

In the second configuration, the critical value of the dimensionless parameter $a$ is determined, with the magnetic charge fixed at $q = 7.5 M$, and the CG parameter and cosmological constant set to $\gamma = 0.5 M^{-1}$ and $\Lambda=0.15 M^{-2}$, respectively. Here, we find $a_c = 3.061$. Fig. \ref{fig:acrit_I} shows the metric function \eqref{eq:A(r)_I} plotted under these considerations for three values of $a$: $a > a_c$, $a = a_c$, and $a < a_c$.
The resulting configuration is similar to the one illustrated in Fig. \ref{fig:qcrit_I}, evidencing a pair of degenerate horizons at $a=a_c$. However, deviations from the critical value of $a$ affect the geometry inversely. Notably, when $a<a_c$ the geometry is a horizonless configuration of a $(-,+,-,-)$ type (similarly to the $q>q_c$ case), whereas the $a>a_c$ geometry contains an inner event horizon and a cosmological horizon (similarly to the $q<q_c$ case). In contrast with the magnetic charge, varying the value of $a$ has a more noticeable effect at $r \sim M$, determining whether the geometry contains one or two maxima.

The previous analyses demonstrate that the geometrical structure of the spacetime in \eqref{eq:A(r)_I} depends on a sensitive interplay between the $a$, $q$, $\gamma$ and $\Lambda$ parameters. The number of horizons is sensitive to the values of $a$ and $q$, though the way in which the metric \eqref{eq:A(r)_I} is affected by each of these parameters is different. Conversely, at large distances the contributions of the linear $\gamma$ term and the $\Lambda$ quadratic term dominate the configuration, with the latter defining the asymptotic nature. Due to the mathematical formulation of the metric function \eqref{eq:A(r)_I}, obtaining an analytical expression for the location of an event horizon is not possible. However, our results also suggest certain values of $a$, $q$, $\Lambda$ and $\gamma$ could lead to horizonless configurations of a $(-,+,-,-)$ type, similarly to a Kottler (Schwarzschild-de Sitter) BH with $\Lambda > 1/9 M^{-2}$ \cite{kottler:1918phy}. 

\begin{figure}[ht!]
    \includegraphics[width=\columnwidth]{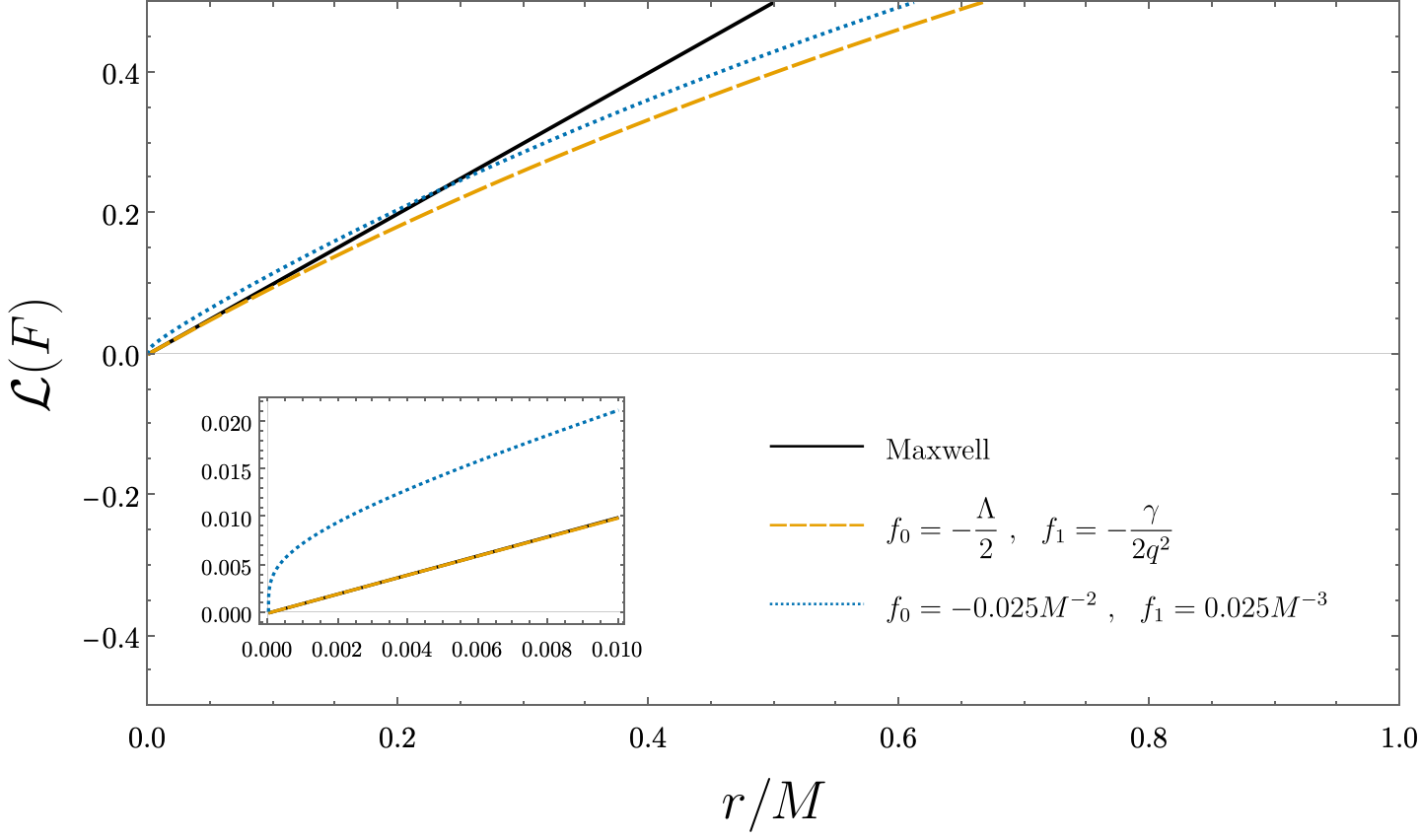}
    \caption{The NLED Lagrangian density ${\cal L}(F)$, as described by Eq. \eqref{eq:L_I}, for two scenarios: $f_0 = -\frac{\Lambda}{2}$ and $f_1 = \frac{\gamma}{2 q^2}$ (dashed gold line); $f_0 = -0.025 M^{-2}$ and $f_1 = 0.025 M^{-3}$ (dotted blue line). Model parameters: $a=0.5$, $q=M$, $\gamma = 0.02 M^{-1}$, $\Lambda=0.05 M^{-2}$.}
    \label{fig:L_I}
\end{figure}

We now proceed to the general expression of ${\cal L}(F)$ for this model. This is achieved by substituting Eq. \eqref{eq:A(r)_I} into Eq. \eqref{eq:Lgeneral}, while resorting to Eq. \eqref{eq:F} to write the NLED Lagrangian density as function of the scalar $F$ as
\begin{equation}
    {\cal L}(F)= \frac{F}{1+aF} + \sqrt[4]{\frac{2 F}{q^2}} \left(2 f_1 q^2-\gamma \right)+f_0+\frac{\Lambda}{2}.\label{eq:L_I}
\end{equation}
The first term of Eq. \eqref{eq:L_I} matches with the initially proposed NLED Lagrangian, while its remaining terms include the contributions stemming from CG. In this form, this equation remains nonlinear in the weak-field approximation due to the presence of a $\sqrt[4]{F}$ term. However, when the integration constants obey the conditions $f_0 = -\frac{\Lambda}{2}$ and $f_1 = \frac{\gamma}{2 q^2}$, Eq. \eqref{eq:L_I} satisfies both the Maxwell and the first-order QED weak field limits \cite{Bokulic:2023afx} as ${\cal L}(F) \sim F$ and ${\cal L}(F) \sim F - a F^2$, respectively, when $F \ll 1$ and $a>0$. As an example of such behavior, the Lagrangian in Eq. \eqref{eq:L_I} is represented in Fig. \ref{fig:L_I}, for two distinct scenarios. In one case, the values of the integration constants $f_0$ and $f_1$ are arbitrarily set to $f_0 = -0.025M^{-2}$ and $f_1 = 0.025M^{-3}$ (dotted blue line), while in the second case (dashed gold line) they obey the previous conditions. In both instances, the model parameters are set to $a=0.5$, $q=M$, $\gamma = 0.02 M^{-1}$, $\Lambda=0.05 M^{-2}$.
As can be seen in Fig. \ref{fig:L_I}, the nonlinear nature of ${\cal L}(F)$ is noticeable, particularly when contrasted against the linear case (i.e., ${\cal L}(F) = F$). The differences between each case are emphasized in the subplot of Fig. \ref{fig:L_I} where, for small values of $F$, the Lagrangian with the arbitrarily defined values of $f_0$ and $f_1$ deviates from the classical Maxwell electrodynamics. 

\begin{figure}[ht!]
    \includegraphics[width=\columnwidth]{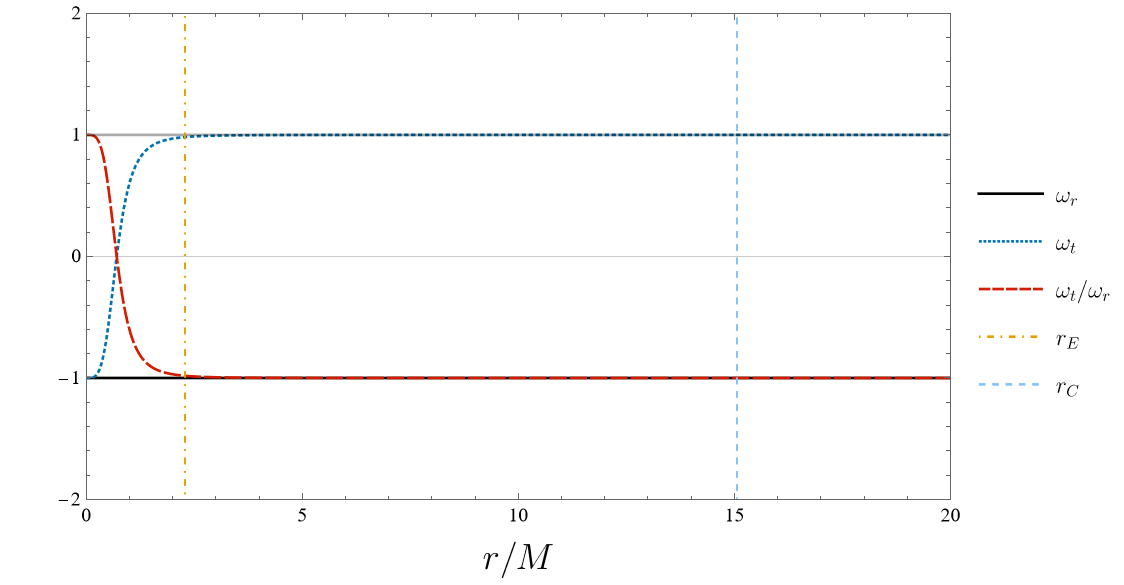}
    \caption{The $\omega_r, \omega_t$ and $\omega_t/\omega_r$ functions, as described by Eq. \eqref{eq:omegas}, computed for Model I. Model parameters: $a=0.5$, $q=M$, $\gamma = 0.2 M^{-1}$, $\Lambda=0.05 M^{-2}$. $r_E$ and $r_C$ denote the event horizon and the cosmological horizon, respectively.}
    \label{fig:ECs_I}
\end{figure}

Lastly, we benchmark this model against the standard energy conditions. These conditions, known as the null (NEC), weak (WEC), strong (SEC), and dominant (DEC) energy conditions, express classical physical behavior—such as the non-negativity of energy density and the causal propagation of energy—and are a powerful tool for assessing the classical plausibility of a given model. They are formulated as \cite{Hawking:1973uf, Visser:1995cc, Lobo:2020ffi}:
\begin{align}
   & NEC_{1,2}=WEC_{1,2}=SEC_{1,2}\Longleftrightarrow \rho + p_{r,t} \geq 0,\label{eq:EC_1}\\
   & SEC_3 \Longleftrightarrow \rho + p_r + 2p_t \geq 0,\label{eq:EC_2}\\
   & DEC_{1,2} \Longleftrightarrow \rho - |p_{r,t}| \geq 0,\label{eq:EC_3}\\
   & DEC_3 = WEC_3 \Longleftrightarrow \rho \geq 0,\label{eq:EC_4}
\end{align}
where $\rho$, $p_r$ and $p_t$ are obtained from the energy-momentum tensor $T^{\mu}_{\phantom{\nu}\nu}=\text{diag}[ \rho, -p_r, -p_t, -p_t ]$. By defining
\begin{equation}\label{eq:omegas}
    \omega_r=\frac{p_r}{\rho}, \qquad \omega_t=\frac{p_t}{\rho}, \qquad \omega_r=\frac{p_t}{p_r},
\end{equation}
we are able to check the energy conditions for each model by analyzing $\omega_{r,t} \geq -1$ \eqref{eq:EC_1}, $\omega_t \geq 0$ \eqref{eq:EC_2}, $|\omega_{r,t}| \leq 1 \eqref{eq:EC_3}$, and $\rho \geq 0$ \eqref{eq:EC_4}. Considering Model I, the $DEC_3$ and $WEC_3$ conditions are verified for all $r$, when $a > 0$. In Fig. \ref{fig:ECs_I}, we examine the remaining conditions for the case of $a=0.5, q=M, \gamma=0.2 M^{-1}$, and $\Lambda=0.05M^{-2}$. The $SEC_3$ condition is the only energy condition that is violated, when $r<\sqrt[4]{\frac{aq^2}{2}}$. In the case considered here, this occurs inside the geometry's event horizon. Otherwise, the model is in agreement with the remaining energy conditions. We note that the case represented in Fig. \ref{fig:ECs_I}, is just one of many possible configurations, and certain combinations of parameter values may lead to geometries where the $SEC_3$ condition is violated outside the event horizon (i.e., $r_E<\sqrt[4]{\frac{aq^2}{2}}$).

\subsection{Model II}\label{ssec:II}

The second spacetime geometry that we analyze is obtained by assuming the following form for Lagrangian density: ${\cal L}(F)=F\left(\frac{1-F}{1+F}\right)$. Likewise, this functional form draws inspiration from another NLED model introduced in \cite{Kruglov:2014hpa}. When substituting the previous expression into equations \eqref{eq:CGfieldeqs1} and \eqref{eq:CGfieldeqs2}, one obtains the following metric function
\begin{align}\label{eq:A(r)_II}
   & A(r)= \, 1 - \frac{2M}{r} + \gamma r - \frac{\Lambda}{3}r^2 - \frac{q^2}{r^2}+q \sqrt[4]{F}\times
\nonumber\\
&\Bigg[ \arctan \left(S^{-}\right) - \arctan \left(S^{+}\right) + \frac{1}{2}\log\left(\frac{S^{+}\sqrt{F} + 1}{S^{-}\sqrt{F} + 1}\right) \Bigg],
\end{align}
where we define $S^{\pm}=1 \pm \sqrt[4]{\frac{4}{F}}$,
as well as utilize Eq. \eqref{eq:F} to simplify the expression. As with the previous model, the limit $q \rightarrow 0$ recovers the CG vacuum solution \eqref{eq:CGvacBH}. 

\begin{figure}[ht!]
    \includegraphics[width=\columnwidth]{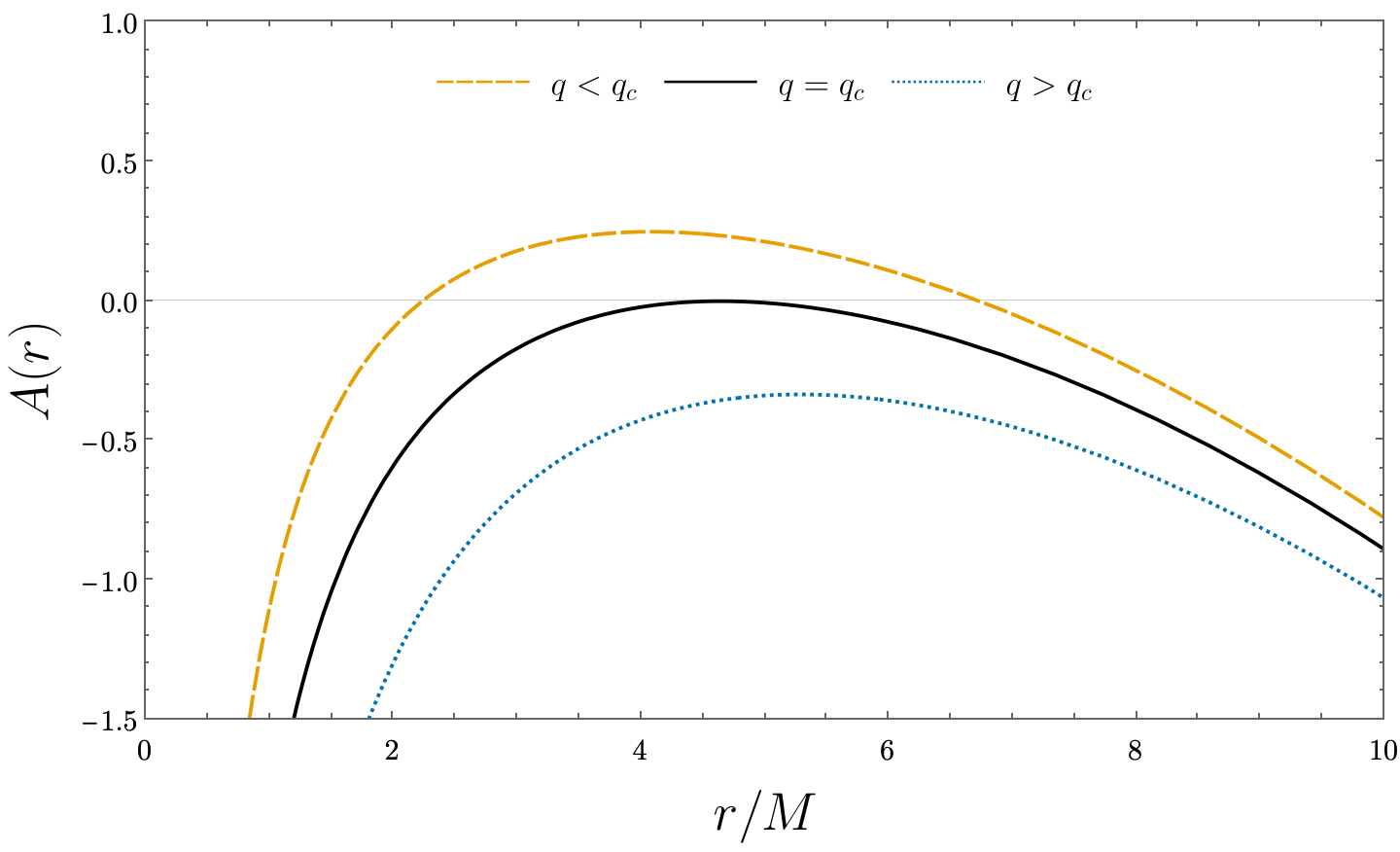}
    \caption{The metric function $A(r)$ of Model II, as described by Eq. \eqref{eq:A(r)_II}, for three distinct charge values: $q < q_c$, $q = q_c$, and $q > q_c$, where $q_c = 0.6412 M$. Model parameters: $\gamma=0.01 M^{-1}$, $\Lambda = 0.05 M^{-2}$.}
    \label{fig:qcrit_II}
\end{figure}

Let us now study the spacetime configurations provided by Eq. \eqref{eq:A(r)_II}, focusing on the critical values of $q$ and $\gamma$ for spacetime configurations where $\Lambda>0$. In the first configuration, the critical value of the magnetic charge $q_c$ is found to be $q_c = 0.6412 M$ when assuming $\gamma =0.01 M^{-1}$ and $\Lambda =0.05 M^{-2}$. Fig. \ref{fig:qcrit_II} illustrates the behavior of the metric function \eqref{eq:A(r)_II} according to the previous assumptions for three distinct charge values: $q > q_c$, $q = q_c$, and $q < q_c$.
Inspecting Fig. \ref{fig:qcrit_II}, one sees that the $q = q_c$ scenario consists of an extreme black hole which contains a degenerate double horizon. Below the critical charge, the geometry possesses an event horizon and a cosmological horizon, whereas above the critical charge, the resulting geometry has no horizons since it is always of a $(-,+,-,-)$ nature. This yields a total of three possible spacetime geometries for the considered parameter values.

\begin{figure}[ht!]
    \includegraphics[width=\columnwidth]{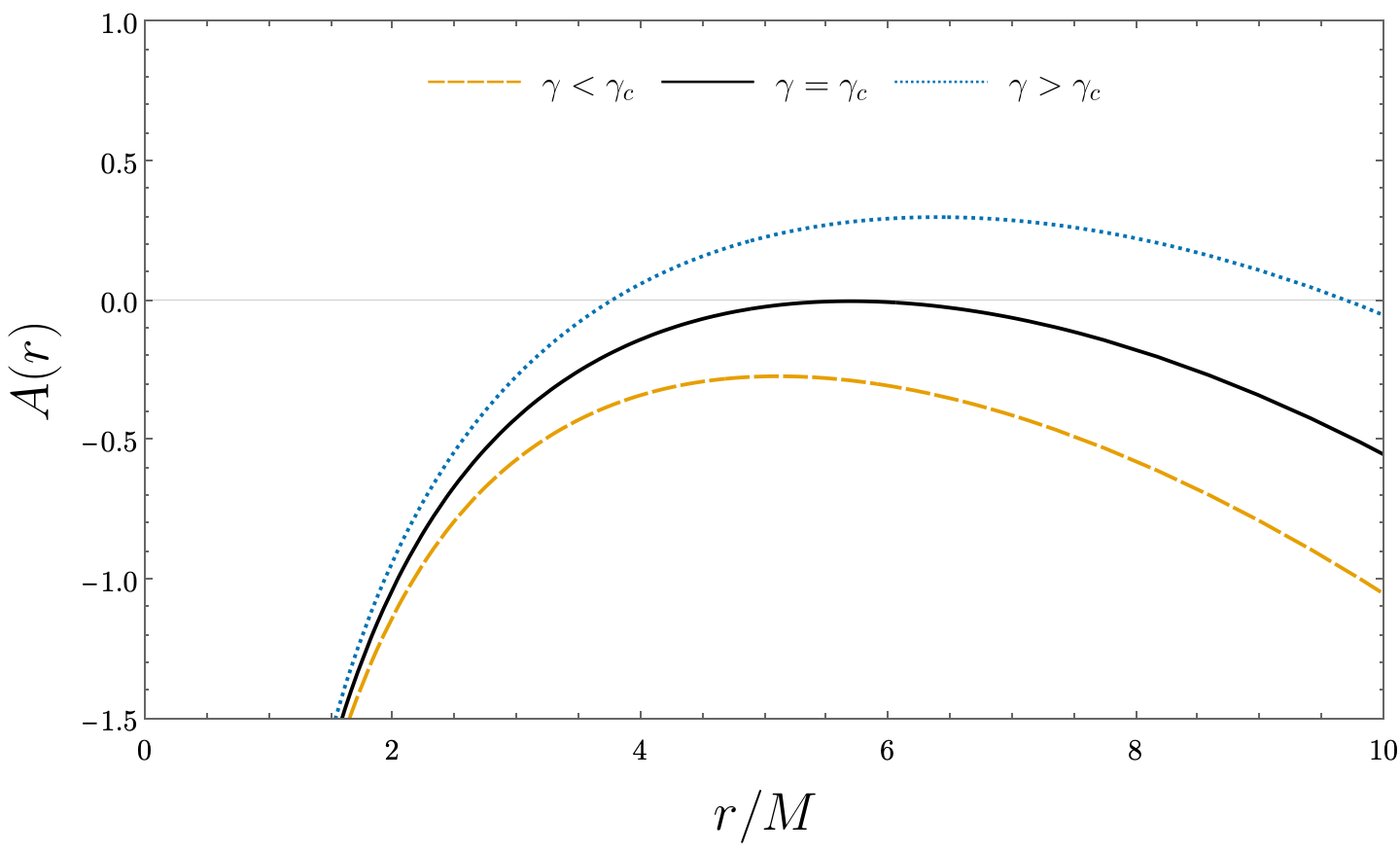}
    \caption{The metric function $A(r)$ of Model II, as described by Eq. \eqref{eq:A(r)_II}, for three distinct values of the CG parameter: $\gamma > \gamma_c$, $\gamma = \gamma_c$, and $\gamma < \gamma_c$, with $\gamma_c = 0.05703M^{-1}$. Model parameters: $q = M$, $\Lambda = 0.05 M^{-2}$.}
    \label{fig:gcrit_II}
\end{figure}

In the second configuration, the critical value of  $\gamma$ is determined when the charge is fixed at $q = M$ and $\Lambda=0.05 M^{-2}$. Here, we find the critical CG parameter is $\gamma_c = 0.05703M^{-1}$. Fig. \ref{fig:gcrit_II} shows the metric function \eqref{eq:A(r)_II} plotted under these considerations for three scenarios: $\gamma > \gamma_c$, $\gamma = \gamma_c$, and $\gamma < \gamma_c$. 
Qualitatively, the cases represented in Fig. \ref{fig:gcrit_II} are identical to those studied in the first configuration, consisting of three possible spacetime structures. Notably, one can identify an extreme black hole configuration when $\gamma = \gamma_c$, where the geometry contains a degenerate double horizon at $r \sim 6M$. Additionally, one geometry possesses an event horizon and a cosmological horizon, whereas the remaining one has no horizons, and it is always of a $(-,+,-,-)$ nature. Deviating from $\gamma_c$ now has an opposite effect to the one caused by deviations from $q_c$, as seen in the first configuration. For example, where the scenario with two horizons previously occurred when $q < q_c$, it now occurs for $\gamma > \gamma_c$. 

Unlike model Model I, however, we note the absence of any significant relations between the cosmological constant and the remaining parameters. Indeed, though purely spacelike configurations are possible via specific values of $q$, $\gamma$, and $\Lambda$, we found no combination of the latter two that could result in a spacelike geometry regardless of the magnetic charge value. 

\begin{figure}[ht!]
    \includegraphics[width=\columnwidth]{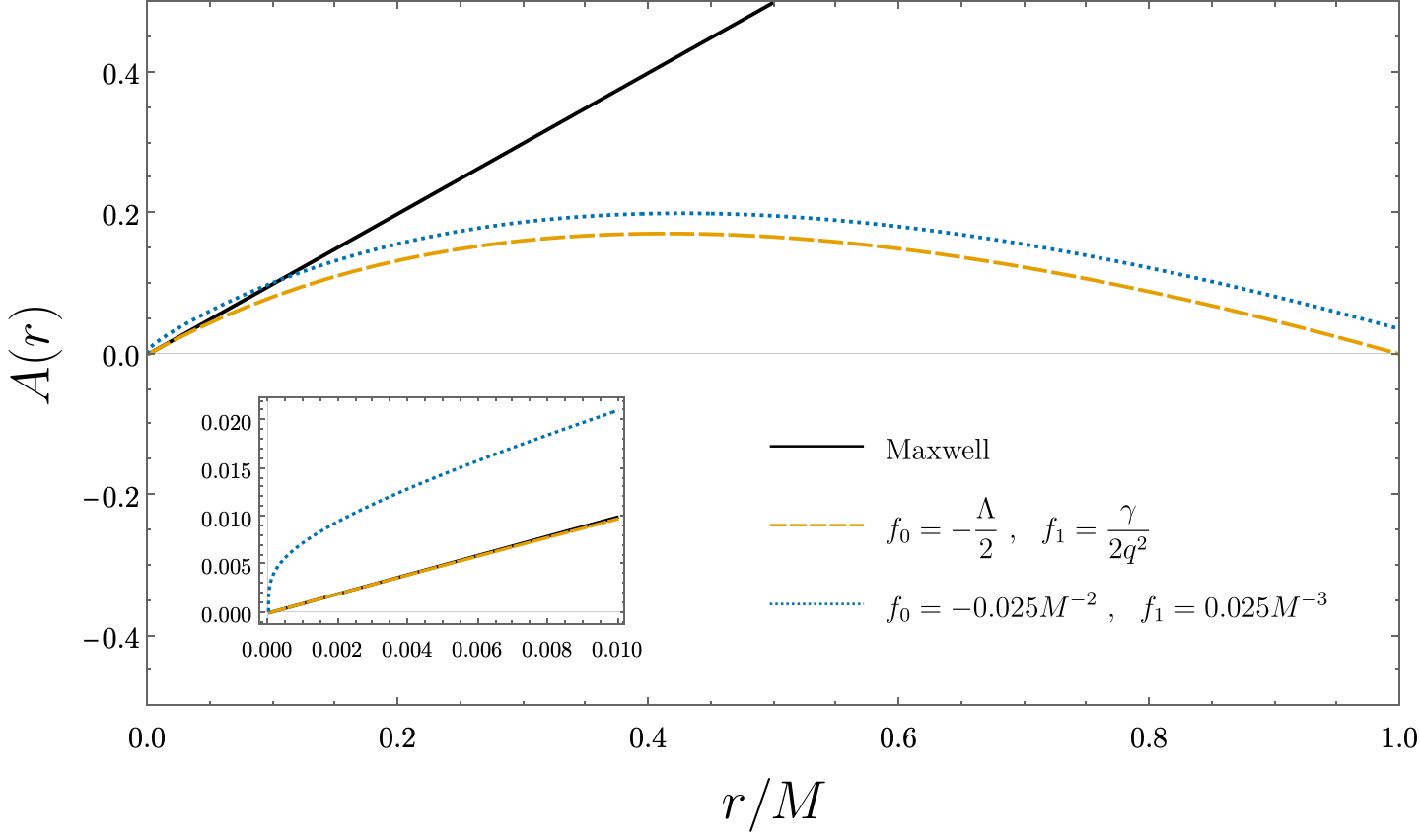}
    \caption{The NLED Lagrangian density ${\cal L}(F)$, as described by Eq. \eqref{eq:L_II}, for two scenarios: $f_0 = -\frac{\Lambda}{2}$ and $f_1 = \frac{\gamma}{2 q^2}$ (dashed gold line); $f_0 = -0.025 M^{-2}$ and $f_1 = 0.025 M^{-3}$ (dotted blue line). Model parameters: $q=M$, $\gamma = 0.02 M^{-1}$, $\Lambda=0.05 M^{-2}$.}
    \label{fig:L_II}
\end{figure}

We now move towards this model's general form of the Lagrangian ${\cal L}(F)$. Substituting Eq. \eqref{eq:A(r)_II} into Eq. \eqref{eq:Lgeneral}, in combination with Eq. \eqref{eq:F}, provides the following expression for NLED Lagrangian density
\begin{equation}
    {\cal L}(F)= F\left(\frac{1-F}{1+F} \right)+ \sqrt[4]{\frac{2 F}{q^2}} \left(2 f_1 q^2-\gamma \right)+f_0+\frac{\Lambda}{2},\label{eq:L_II}
\end{equation}
whose first term corresponds to the chosen NLED Lagrangian for this model. The following $\sqrt[4]{F}$ terms render Eq. \eqref{eq:L_II} nonlinear in the weak-field approximation; a feature that may be circumvented via a suitable choice of values for the integration constants $f_0 = -\frac{\Lambda}{2}$ and $f_1 = \frac{\gamma}{2 q^2}$. 
In such instances, Eq. \eqref{eq:L_II} also recovers the classical electrodynamics regime for $F \ll 1$, while mimicking Euler–Heisenberg-like corrections \cite{Yajima:2000kw}. Following the same procedure as Model I, we study the behavior of the Lagrangian \eqref{eq:L_II} in  Fig.\ref{fig:L_II}, for two scenarios: in one case, the values of the integration constants $f_0$ and $f_1$ are set to $f_0 = -0.025 M^{-2}$ and $f_1 = 0.025 M^{-3}$ (dotted blue line); in the second case, the integration constants $f_0$ and $f_1$ obey $f_0 = -\frac{\Lambda}{2}$ and $f_1 = \frac{\gamma}{2 q^2}$ (dashed gold line). In both instances, the model parameters are set to $q=M$, $\gamma = 0.02 M^{-1}$, $\Lambda=0.05 M^{-2}$.
Observing Fig. \ref{fig:L_II}, the nonlinear character of ${\cal L}(F)$ stands out when compared with the linear (i.e., Maxwell) electrodynamics Lagrangian, particularly as the field strength $F$ increases. The zoom-in in Fig. \ref{fig:L_II}'s subplot highlights the difference between the two considered cases, where only the scenario with arbitrarily defined values of $f_0$ and $f_1$ deviates from the classical Maxwell electrodynamics.

\begin{figure}[ht!]
    \includegraphics[width=\columnwidth]{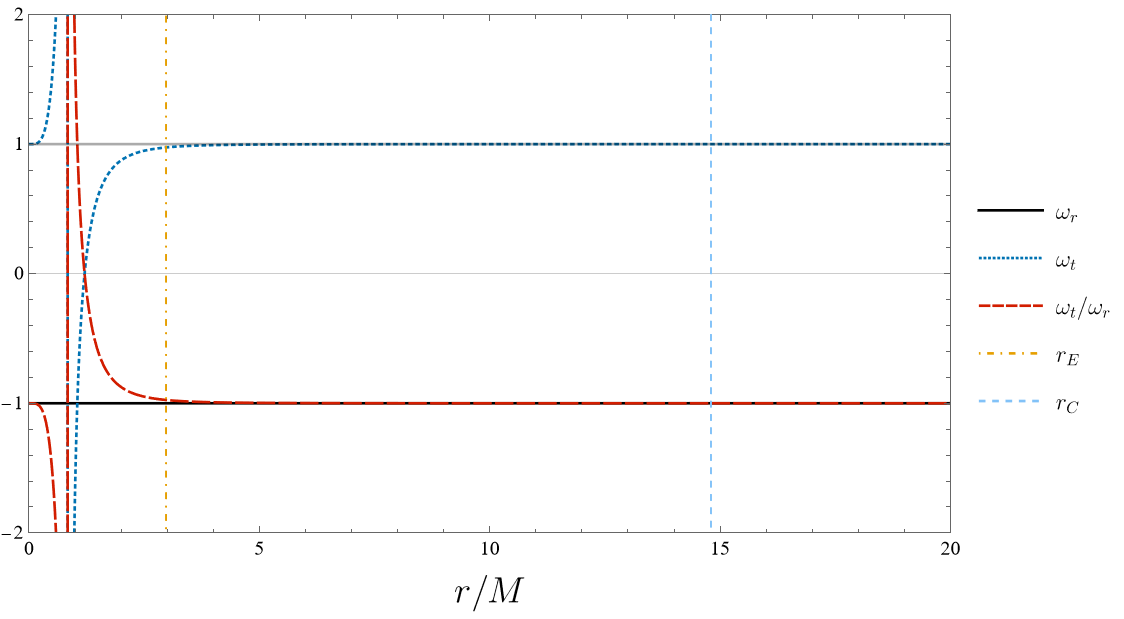}
    \caption{The $\omega_r, \omega_t$ and $\omega_t/\omega_r$ functions as described by Eq. \eqref{eq:omegas}, computed for Model II. $r_E$ and $r_C$ denote the event horizon and the cosmological horizon, respectively. Model parameters: $q=M$, $\gamma = 0.2 M^{-1}$, $\Lambda=0.05 M^{-2}$.}
    \label{fig:ECs_II}
\end{figure}

Next, we test Model II against the energy conditions given by Eqs. \eqref{eq:EC_1} to \eqref{eq:EC_4}. For this model, the $DEC_3$ and $WEC_3$ conditions are verified for all $r\geq \sqrt[4]{\frac{q^2}{2}}$. If $r_E < \sqrt[4]{\frac{q^2}{2}}$ the $DEC_3$ and $WEC_3$ are violated outside the event horizon. Moreover, we find that: the $SEC_3$ is violated when $r < \sqrt[4]{\frac{q^2}{2}(2+\sqrt{5})}$; $NEC_{2},WEC_{2},SEC_{2}$ and $DEC_2$ are violated when $r < \sqrt[4]{\frac{q^2}{2}(1+\sqrt{2})}$; $DEC_1$ is violated when $r < \sqrt[4]{\frac{q^2}{2}}$. In Fig. \ref{fig:ECs_II}, where we examine the case of $ q=M, \gamma=0.2 M^{-1}$, and $\Lambda=0.05M^{-2}$, all the energy condition violations take place at some point inside the event horizon. Outside the event horizon, the model is in compliance with all the energy conditions.

\subsection{Model III}\label{ssec:III}

The third spacetime model we present is obtained from a Lagrangian density  described by a general power-law function ${\cal L}(F)=F+aF^k$, which introduces the coupling constant $a$ and the exponent $k$. This type of power law model, in addition to including the quadratic order terms of the Euler–Heisenberg effective Lagrangian (at $k=2$), can also include higher-order contributions at $k>2$ which may arise from higher-loop corrections \cite{Dunne:2004nc}. The particular case of $a=-1$ and $k=1$ returns the CG vacuum solution \eqref{eq:CGvacBH}, since the matter content described by the power-law Lagrangian ${\cal L}(F)=F+aF^k=0$ vanishes. The metric function that satisfies Eqs. \eqref{eq:CGfieldeqs1} and \eqref{eq:CGfieldeqs2}, holds the following expression
\begin{equation}\label{eq:A(r)_III}
A(r)= 1 -\frac{2 M}{r} +\frac{q^2}{r^2} +\gamma r -\frac{\Lambda}{3}r^2 +\frac{a2^{1-k}q^{2k}}{4k-3} r^{2-4k}.
\end{equation}
In contrast with the models presented in \ref{ssec:I} and \ref{ssec:II}, Eq. \eqref{eq:A(r)_III} contains the two additional free parameters, $a$ and $k$, stemming from its parent ${\cal L}(F)$ function. The parameter $k$ describes a dimensionless quantity that is present the last term's denominator as well as in the exponent of $r$. This has a number of consequences: from the denominator, the value $k=\frac{3}{4}$ is outright excluded (i.e., $k \in \mathbb{R}/\{\frac{3}{4}\}$); the behavior of the spacetime configuration is extremely sensitive to values of $k \sim 3/4$; specific values of $k$ represent thresholds with regard to this last term's contribution to the spacetime geometry, as well as determine the dimension of the constant $a$ (i.e. $a$ has geometrized units of $M^{2k-2}$).

\begin{figure}[ht!]
    \includegraphics[width=\columnwidth]{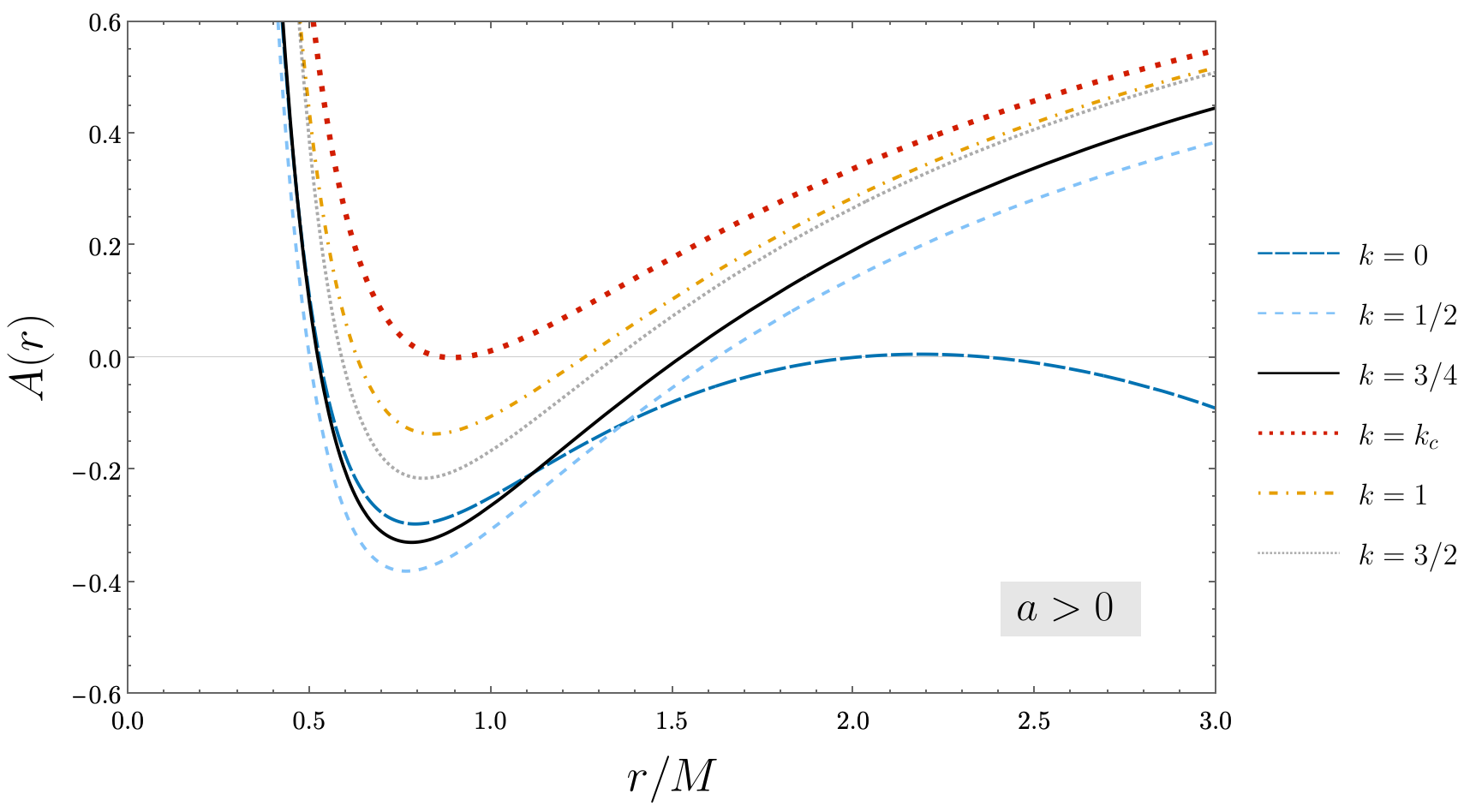}
    \caption{The metric function $A(r)$ of Model III, as described by Eq. \eqref{eq:A(r)_III} and Eq. \eqref{eq:A(r)_III_3/4}, for $k \in \{ 0, \frac{1}{2}, \frac{3}{4}, k_c, 1, \frac{3}{2} \}$, where $k_c=0.8633$. Model parameters: $q=0.88M$, $\gamma=0.05 M^{-1}$, $\Lambda = 0.02 M^{-2}$, $a=0.1M^{2k-2}$.}
    \label{fig:Kcrit_III}
\end{figure}

Let us focus on the particular case of $k=\frac{3}{4}$. As previously mentioned, this value is not permitted, as it leads to an undefined expression in the last term of Eq. \eqref{eq:A(r)_III}. To circumvent this issue, one may instead search for a solution of Eqs. \eqref{eq:CGfieldeqs1} and \eqref{eq:CGfieldeqs2}, by considering the particular case of ${\cal L}(F)=F+aF^{\frac{3}{4}}$. In doing so, the following metric function is obtained
\begin{align}\label{eq:A(r)_III_3/4}
   A(r)= & \, 1-\frac{2 M}{r} + \frac{q^2}{r^2} +\gamma  r -\frac{\Lambda}{3} r^2 
\nonumber\\
& -2a\left(\frac{5}{6}+\log{\frac{r}{M_0}}\right)F^{\frac{3}{4}}r^2,
\end{align}
containing only the parameters $q,\gamma,\Lambda$ and $a$ as free parameters, with the latter possessing dimensions of $M^{-1/2}$. Fig. \ref{fig:Kcrit_III} displays the metric functions \eqref{eq:A(r)_III} and \eqref{eq:A(r)_III_3/4} with respect to $r$, for multiple values\footnote{We disregard values in the interval $k<0$, since in these cases the resulting geometry is singular in the infinite future.} of the exponent $k$ (i.e., $k \in \{ 0, \frac{1}{2}, \frac{3}{4}, k_c, 1, \frac{3}{2}  \}$), with $k_c$ denoting the critical value of the exponent $k$. We assume $q=0.88M$, $\gamma=0.05 M^{-1}$, $\Lambda=0.02 M^{-2}$, $a=0.1M^{2k-2}$ and find the critical exponent to be $k_c=0.8633$.
As can be seen in Fig. \ref{fig:Kcrit_III}, several possible geometries emerge depending on the value of $k$. When $k=0$, the resulting geometry contains three event horizons along with the additional cosmological horizon. In this case, the last term of Eq. \eqref{eq:A(r)_III} is proportional to $r^2$, which results in $a$ having dimensions of $M^{-2}$ and playing the role of an effective cosmological constant. When $k=\frac{1}{2}$, the geometry contains three horizons (i.e., two even horizons and a cosmological one). However, the last term of Eq. \eqref{eq:A(r)_III} is proportional to $-2a$, which results in a dimensionless $a$. Additionally, the local minimum $A(r)$ is now more pronounced compared to the $k=0$ geometry. The $k=\frac{3}{4}$ geometry also contains three horizons (under the assumed parameter values). In this scenario, the last terms of Eq. \eqref{eq:A(r)_III_3/4} are proportional to $r^{-1}$ and $r^{-1}\log{r}$, which results in $a$ having dimensions of $M$. More importantly, this represents a departure from the previous trend, as the resulting geometry is akin to an asymptote from which the metric \eqref{eq:A(r)_III} departs as one considers values of $k \rightarrow \frac{3}{4}$. For geometries that $k\lesssim \frac{3}{4}$ the local minimum diverges to $-\infty$, whereas geometries with $k\gtrsim \frac{3}{4}$ approach the "asymptote" from above. This suggests that, in the absence of the cosmological horizon (i.e., $\Lambda=0$), naked singularity configurations are possible in the interval $\frac{3}{4} < k < k_c$, within the considered parameter values of $q$, $\gamma$, and $a$. When $k>k_c$, the resulting configurations always contain three event horizons, as can be seen from the $k=1$ and $k=\frac{3}{2}$ cases.

\begin{figure}[ht!]
    \includegraphics[width=\columnwidth]{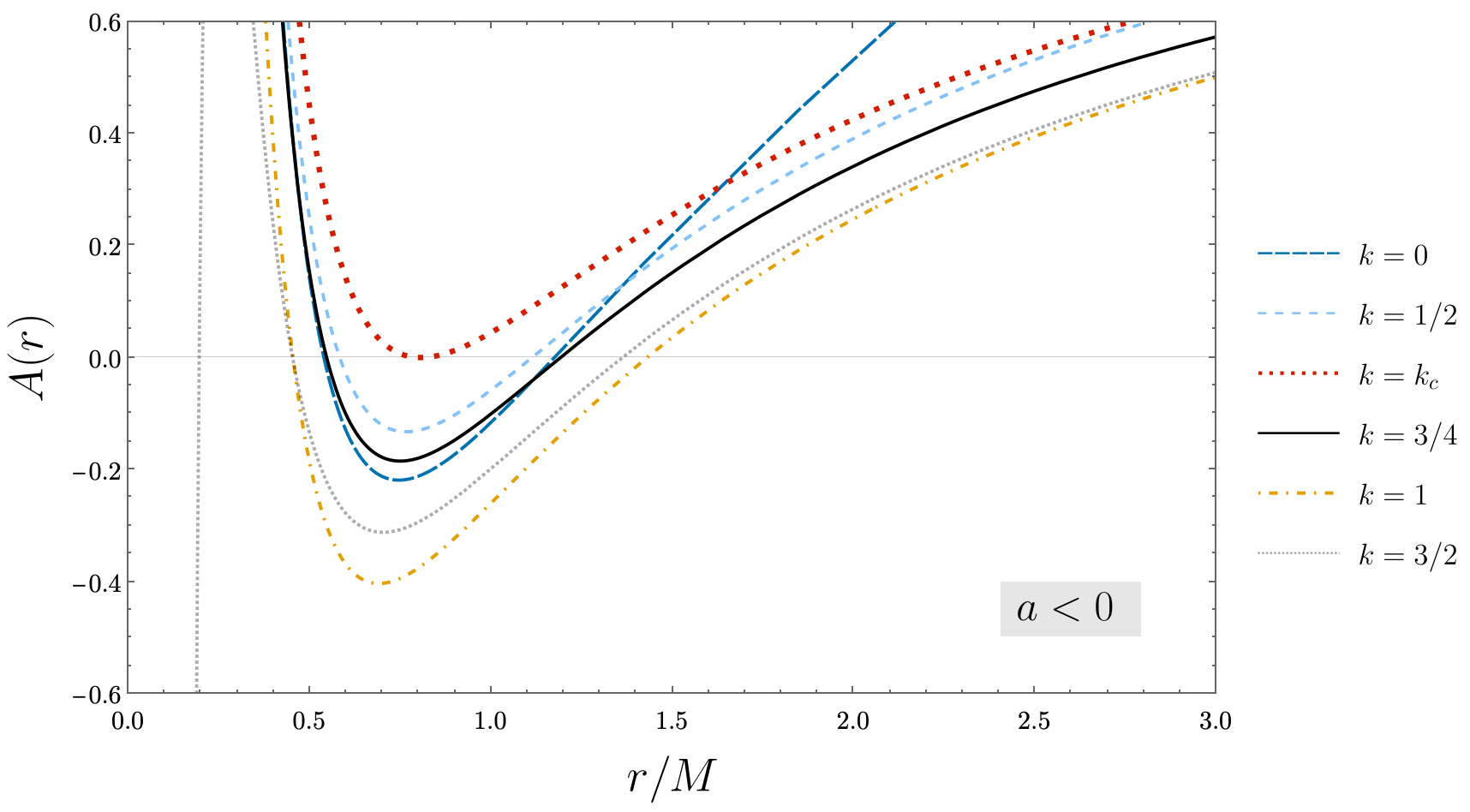}
    \caption{The metric function $A(r)$ of Model III, as described by Eq. \eqref{eq:A(r)_III} and Eq. \eqref{eq:A(r)_III_3/4}, for $k \in \{ 0, \frac{1}{2}, \frac{3}{4}, k_c, 1, \frac{3}{2} \}$, with $k_c=0.6286$. Model parameters: $q=0.88M$, $\gamma=0.05 M^{-1}$, $\Lambda = 0.02 M^{-2}$, $a=-0.1M^{2k-2}$.}
    \label{fig:Kcrit_III_neg}
\end{figure}

Similarly, Fig. \ref{fig:Kcrit_III_neg} displays the metric functions \eqref{eq:A(r)_III} and \eqref{eq:A(r)_III_3/4} with respect to $r$, for multiple values of the exponent $k$ (i.e., $k \in \{ 0, \frac{1}{2}, \frac{3}{4}, k_c, 1, \frac{3}{2}  \}$), but now in the context of $a < 0$. We consider the same parameter values of $q, \gamma$ and $\Lambda$ as before, with the exception of $a$ which now assumes the value $a=-0.1M^{2k-2}$. In such a context, we find the critical exponent to be $k_c=0.6286$. From Fig. \ref{fig:Kcrit_III_neg}, it is possible to observe how the change in the sign of $a$ affects the various configurations represented therein.
When $k=0$, the contribution of $a$ as an effective cosmological constant overcomes that of $\Lambda$, resulting in a geometry with two event horizons and no cosmological horizon. As the value of the exponent $k$ increases to $k\lesssim \frac{3}{4}$ the geometries' local minimum diverges to $\infty$, suggesting once more the existence of naked singularity configurations when $\Lambda=0$. The $k=\frac{3}{4}$ geometry contains three horizons, with a similar shape to its $a>0$ counterpart. For geometries with $k\gtrsim \frac{3}{4}$, the local minimum approaches the $k=\frac{3}{4}$ case from below. When $k>1$, the resulting configurations contain three event horizons along with a cosmological horizon. This occurs due to the presence of negative terms of $r$ with an exponent smaller than $2$, whose contribution becomes dominant when $r \ll M$.

\begin{figure}[ht!]
    \includegraphics[width=\columnwidth]{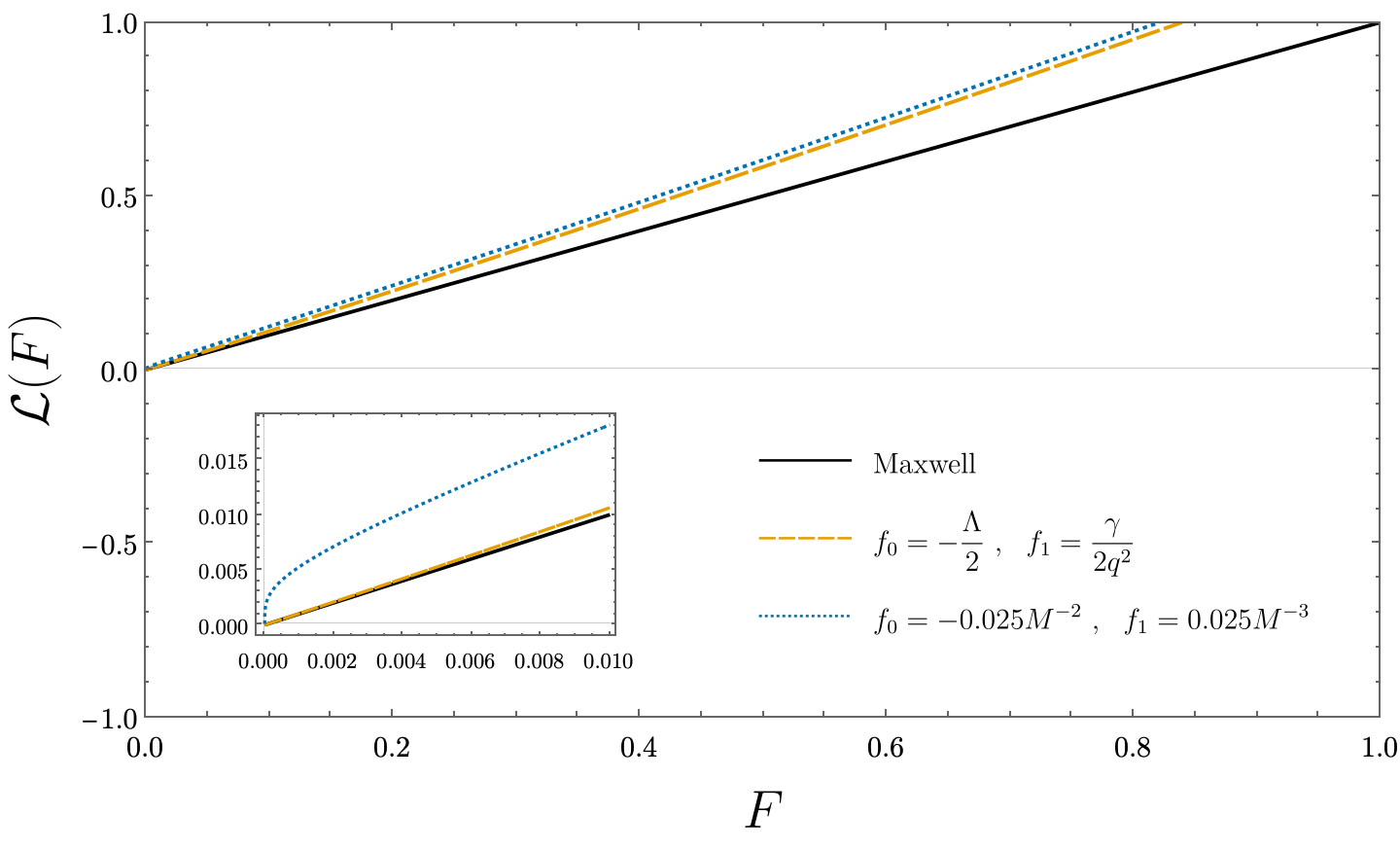}
    \caption{The NLED Lagrangian density ${\cal L}(F)$, as described by Eq. \eqref{eq:L_III}, for two scenarios: $f_0 = -\frac{\Lambda}{2}$ and $f_1 = \frac{\gamma}{2 q^2}$ (dashed gold line); $f_0 = -0.025 M^{-2}$ and $f_1 = 0.025 M^{-3}$ (dotted blue line). Model parameters: $q=M$, $\gamma = 0.02 M^{-1}$, $\Lambda=0.05 M^{-2}$, $a=0.2M^{\frac{1}{2}}$, $k=1.25$.}
    \label{fig:L_III}
\end{figure}

Proceeding towards the general expression of ${\cal L}(F)$ for this model, substituting Eq. \eqref{eq:A(r)_III} into Eq. \eqref{eq:Lgeneral} yields the following NLED Lagrangian density
\begin{equation}
    {\cal L}(F)= F + aF^k + \sqrt[4]{\frac{2 F}{q^2}} \left(2 f_1 q^2-\gamma \right)+f_0+\frac{\Lambda}{2}.\label{eq:L_III}
\end{equation}
In this form, ${\cal L }(F)$ is not consistent with Maxwell electrodynamics in the weak field limit, due to the presence of a $\sqrt[4]{F}$ term along with the term defined by the exponent $k$. In the $f_0 = -\frac{\Lambda}{2}$ and $f_1 = \frac{\gamma}{2 q^2}$ conditions, the Maxwell weak field limit is recovered as long as $k \geq 1$. Thus, for the particular case of $k=\frac{3}{4}$, it is not possible to recover Maxwell electrodynamics in the weak field limit. Note that it is possible to fully recover the Maxwell (linear) electrodynamics for $k=\frac{1}{4}$ if integration constants obey $f_1 = \frac{\gamma-a(q^2/2)^{1/4}}{2 q^2}$ and $f_0 = -\frac{\Lambda}{2}$. Furthermore,  when considering $k=2$ and $a=-16 \kappa$, Eq. \eqref{eq:L_III} satisfies the first order of QED corrections in the weak field regime \cite{Bokulic:2023afx} (i.e. ${\cal L}(F) \sim F - \kappa(4F)^2$), where the parameter $\kappa$ is related to the fine-structure constant and the electron mass. In Fig.~\ref{fig:L_III}, we compare two cases of Lagrangian \eqref{eq:L_III} considering $k=\frac{5}{4}$: in the first case $f_0 = -\frac{\Lambda}{2}$ and $f_1 = \frac{\gamma}{2 q^2}$ (dashed gold line); in the second case $f_0 = -0.025 M^{-2}$ and $f_1 = 0.025 M^{-3}$ (dotted blue line). The model parameters are set to $q=M$, $\gamma = 0.02 M^{-1}$, $\Lambda=0.05 M^{-2}$.  
As can be seen in Fig. \ref{fig:L_III}, the considered Lagrangian density (i.e., with $k=\frac{5}{4}$) is nonlinear, displaying greater asymptotic growth than its linear counterpart. Examining the subplot in Fig. \ref{fig:L_III} reveals the difference between the two nonlinear cases in the weak field regime. Notably, ${\cal L}(F)$ approximates classical electrodynamics only when the constants $f_0$ and $f_1$ obey the previously discussed conditions.

\begin{figure}[ht!]
    \includegraphics[width=\columnwidth]{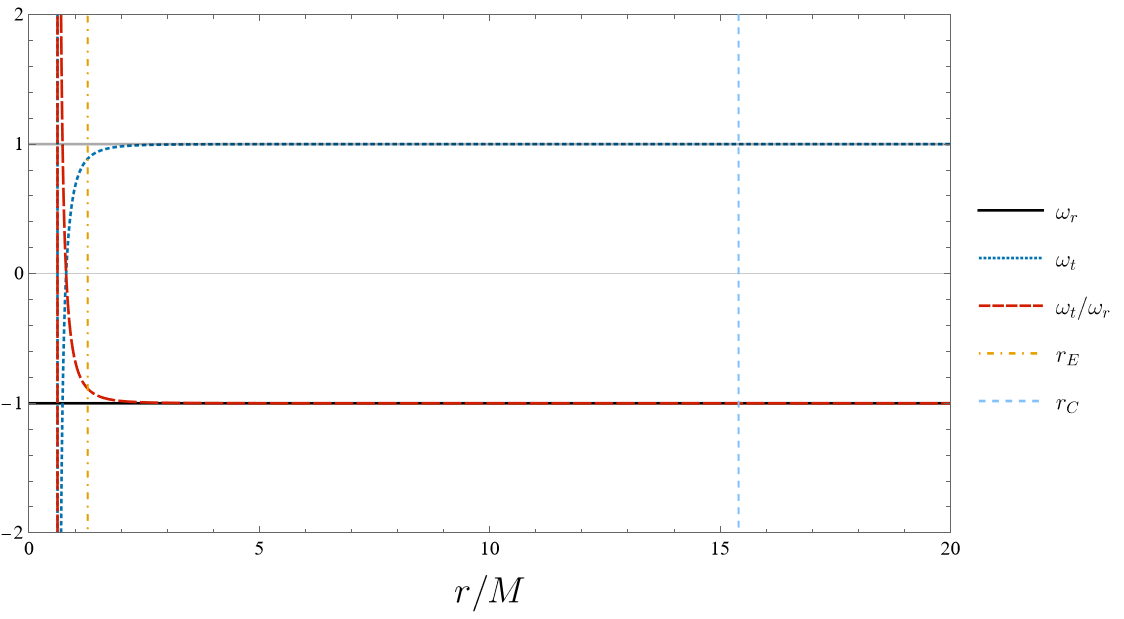}
    \caption{The $\omega_r, \omega_t$ and $\omega_t/\omega_r$ functions as described by Eq. \eqref{eq:omegas}, computed for Model III. $r_E$ and $r_C$ denote the event horizon and the cosmological horizon, respectively. Model parameters: $a=-0.5$, $q=0.75M$, $k=2$, $\gamma = 0.2 M^{-1}$, $\Lambda=0.05 M^{-2}$.}
    \label{fig:ECs_III}
\end{figure}

We conclude this section by testing Model III against the energy conditions given by Eqs. \eqref{eq:EC_1} to \eqref{eq:EC_4}. We restrict our analysis to $a<0$ and $k\geq 1$, covering the range of values at which the QED weak field approximation is satisfied. In this regard, the $DEC_{1,3}$ and $WEC_3$ conditions are respected for all $r\geq \sqrt[4]{\frac{q^2}{2}|a|^{\frac{1}{k-1}}}$. The $NEC_{2},WEC_{2},SEC_{2}$ and $DEC_2$ are verified for all $r\geq \sqrt[4]{\frac{q^2}{2}(k|a|)^{\frac{1}{k-1}}}$. Finally, the $SEC_{3}$ is satisfied when $r\geq \sqrt[4]{\frac{q^2}{2}((2k-1)|a|)^{\frac{1}{k-1}}}$. At $k\gg1$, all of the above converge to $r \sim \sqrt[4]{\frac{q^2}{2}}$ regardless of the value of the parameter $a$. In the case of $k=1$, the energy conditions are fully satisfied when $a \neq -1$. Fig. \ref{fig:ECs_III} depicts the case of $ a=-0.5, \  q=0.75M, \  k=2, \gamma=0.2 M^{-1}$, and $\Lambda=0.05M^{-2}$, where it is possible to observe that all the energy condition violations occur exclusively below the geometry's event horizon. 

\section{Black hole shadows}\label{sec:BHshadows}

In black hole literature, the term \textit{shadow} refers to the dark region in the image plane of an external observer, cast by a black hole that is seen against a bright background, corresponding to the gravitationally lensed projection of a black hole's unstable photon region \cite{Falcke:1999pj, Bronzwaer:2021lzo, Perlick:2021aok}. This unique visual signature encodes information from its caster, since its size and shape are sensitive to the properties of the underlying spacetime geometry \cite{Cunha:2018acu, Chen:2022scf}. In what follows, we provide a brief overview of the calculation of the shadow radius of a generic static and spherically symmetric spacetime (see \cite{Perlick:2021aok} for a review on analytical calculations of the shadow). In such cases, null geodesics satisfy
\begin{equation}
    \left(\frac{dr}{d\phi} \right)^2=\frac{C(r)}{A(r)B(r)}\left( \frac{1}{b^2}-\frac{A(r)}{C(r)}\right),\label{eq:photon_motion}
\end{equation}
where $b$ denotes the geodesic trajectory impact parameter. Consider a static observer located at some distance $r_O$ from a black hole with mass $M$. A light ray can be traced backwards from the observer to its origin, at an angle $\alpha$ relative to the radial coordinate $r$ (e.g. see Fig. 5 in Ref. \cite{Perlick:2021aok}), such that $\alpha$ relates to the metric via
\begin{equation}
    \cot\alpha=\sqrt{\frac{B(r)}{C(r)}}\frac{dr}{d\phi}\Bigg|_{r=r_{O}},\label{eq:lightray_angle}
\end{equation}
which, when substituting Eq. \eqref{eq:photon_motion} into Eq. \eqref{eq:lightray_angle} leads to
\begin{equation}
    \sin^2\alpha=b^2 \frac{A(r)}{C(r)}\Bigg|_{r=r_{O}}.\label{eq:lightray_angle_2}
\end{equation}
The boundary of the shadow coincides with the \textit{critical curve}, the theoretical curve in the observer's image plane that comprises light rays which asymptotically approach the region of unstable bound circular geodesics \cite{Gralla:2019xty}. In such instances, $\frac{dr}{d\phi}=0$ and thus Eq. \eqref{eq:photon_motion} implies
\begin{equation}\label{eq:impact_param}
    b^2=\frac{C(r_{ph})}{A(r_{ph})},
\end{equation}
where $r_{ph}$ is the radius of the unstable photon region, that is a solution of \cite{Claudel:2000yi}
\begin{equation}
    \frac{C^{\prime}(r_{ph})}{C(r_{ph})} = \frac{A^{\prime}(r_{ph})}{A(r_{ph})}\,.
    \label{eq:photon_sphere}
\end{equation}

Inserting Eq. \eqref{eq:impact_param} into Eq. \eqref{eq:lightray_angle_2} in the small angle approximation (i.e., valid at $r_O \gg M$) returns the shadow radius
\begin{equation}
r_{sh}=r_{ph}\sqrt{\frac{A(r_{O})}{A(r_{ph})}},\label{eq:shadow_radius}
\end{equation} 

When the spacetime geometry is described by NLED sources, as is the case of the models considered here, the nonlinearities in the electromagnetic fields display non-trivial behavior in the strong gravity regime. In particular, it has been shown that the presence of nonlinear terms in the matter fields cause photons to travel through an "effective" geometry rather than the background geometry \cite{Novello:1999pg}, which ultimately influences the optical appearance of a BH  \cite{Okyay:2021nnh,Waseem:2025qoo}. The previous derivation must therefore be extended to account for the modifications introduced by NLED. Notably, the effective metric of a geometry whose Lagrangian contains a single non-vanishing field invariant (such as that of purely magnetic configurations) is given by \cite{Novello:1999pg}
\begin{equation}
     g_{(e)}^{\mu\nu}={\cal L}_{F}g^{\mu\nu}-{\cal L}_{FF}F_{\sigma}^{\phantom{\sigma}\mu}F^{\sigma\nu},\label{g_efet}
 \end{equation}
where the notation $_{(e)}$ refers to the effective geometry. In turn, the effective line element may be expressed as 
\begin{equation}\label{eq:eff_metric}
ds^2= H(r)(A(r) dt^2 - B(r)dr^2) - h(r)C(r) d\Omega^2,
\end{equation}
where we have defined the functions $h(r)={\cal L}_F$ and $H(r)={\cal L}_{F}+2F{\cal L}_{FF}$ that contain the deviations between the effective and background metric. Taking Eq. \eqref{eq:eff_metric} into account, we introduce the notation $\widetilde{A}(r)=H(r)A(r)$ and $\widetilde{C}(r)=h(r)C(r)$ and repeat the previous derivation to obtain NLED-upgraded formula for the shadow radius of a static and spherically symmetric geometry
\begin{equation}
r_{sh}=\sqrt{\frac{\widetilde{A}(r_{O})}{\widetilde{A}(r_{ph})}\frac{\widetilde{C}(r_{ph})}{h(r_O)}}.\label{eq:shadow_radius_NLED}
\end{equation} 


We now utilize Eq. \eqref{eq:shadow_radius_NLED} to study the parameter dependent shadow radius of Models I, II, and III. This methodology relies on comparing theoretical expectation of each geometry's shadow radius with the EHT's estimates of Sgr A*'s shadow size. Due to the lack of instrumental sensitivity to photons below a certain threshold of the peak intensity, it is not possible to directly observe the shadow radius of Sgr A*. Nonetheless, the EHT collaboration claims it is possible to circumvent this limitation and infer the shadow radius by using the angular size of the surrounding bright emission ring as its proxy, provided the various bias sources are accounted for (the reader is referred to the discussion in \cite{EventHorizonTelescope:2022urf}). Thanks to prior independent measurements of Sgr A*'s mass-to-distance ratio\footnote{The mass $M$ and distance $D$ to Sgr A* were obtained through precise orbital tracking of the S-stars, a group of stars located in the central stellar cluster in the innermost region of the galactic center (see the discussion on this matter in \cite{EventHorizonTelescope:2022urf}).} $M/D$, by the Keck Observatory \cite{Do:2019txf} and the Very Large Telescope Interferometer (VLTI) \cite{GRAVITY:2020gka} instrumentation teams, the EHT collaboration is able to quantify the fractional deviation $\delta$ between the inferred shadow diameter and the predicted shadow diameter of a Schwarzschild black hole, of angular size  $\theta_{sh,Sch}=6\sqrt{3} M/D$, as \cite{EventHorizonTelescope:2022urf}
\begin{equation}\label{eq:fractional_deviation}
    \delta = \frac{r_{sh}}{3\sqrt{3}M} - 1 \ .
\end{equation}
In turn, the fractional deviation can be converted into meaningful bounds on the observed shadow size of Sgr A*'s shadow radius, at 1$\sigma$
\begin{equation} \label{eq:1sigma}
    4.55 \lesssim r_{s}/M \lesssim 5.22 \ ,
\end{equation}
and at $2\sigma$
\begin{equation} \label{eq:2sigma}
    4.21 \lesssim r_{s}/M \lesssim 5.56 \ .
\end{equation}
These are obtained by averaging the fractional deviation computed from the Keck and VLTI $M/D$ and assuming a gaussian distribution of the uncertainties \cite{Vagnozzi:2022moj}. In the ensuing calculations, we equally consider the average value between the Keck and VLTI's distance measurements for the observer's distance which corresponds to $r_O = 4.1\times 10^{10}M$ in geometrized units.

\subsection{Model I}\label{ssec:I_shadow}

\begin{figure}[ht!]
    \includegraphics[width=\columnwidth]{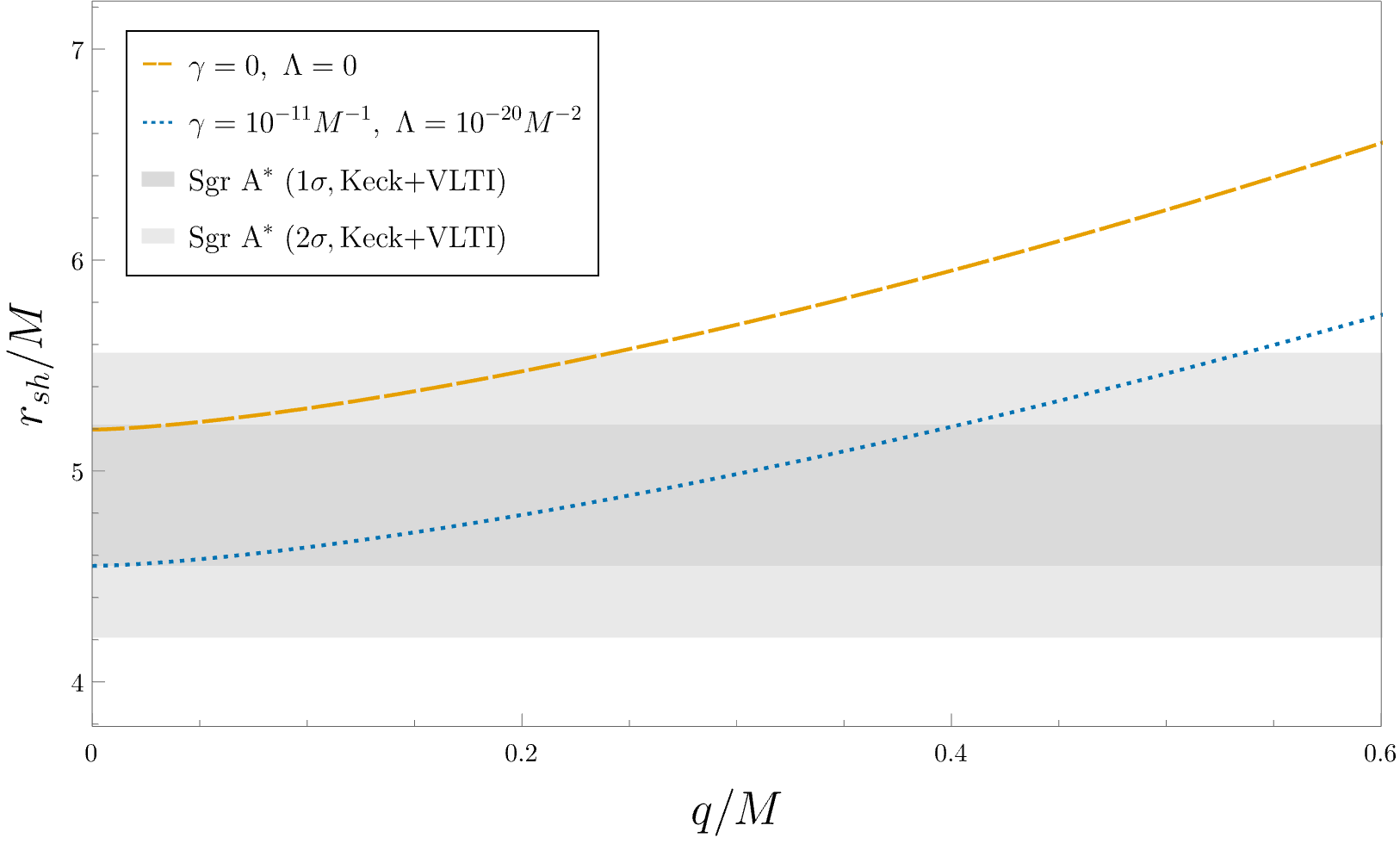}
    \caption{Shadow radius $r_{sh}$ of model I \eqref{eq:A(r)_I} as a function of the magnetic charge $q$ in two scenarios: $a=1$ and $\gamma=\Lambda=0$ (dashed gold line); $a=1$, $\gamma=10^{-11} M^{-1}$ and $\Lambda=10^{-20} M^{-2}$ (dotted blue line). The shaded areas represent the confidence intervals of Sgr A*'s shadow radius at 1$\sigma$ (dark gray) and at 2$\sigma$ (light gray).}
    \label{fig:shadow_I}
\end{figure}

To compute the Model I's shadow radius, let us take its metric functions \eqref{eq:A(r)_I}, $C(r)=r^2$, and its NLED Lagrangian \eqref{eq:L_I}, to obtain the corresponding effective geometry. Due to the already complicated expressions of the background metric function and NLED Lagrangian, this process  results in very cumbersome expression. In particular, it is impossible to solve analytically to retrieve the photon sphere radius and the shadow radius. Thus, we compute these quantities numerically. When performing this correction, the effective geometry is described by an additional fourth parameter, the integration constant $f_1$, due to its presence in the Lagrangian's derivatives. Since the Maxwell Electrodynamics in the weak field is recovered, when $f_1=\frac{\gamma}{2 q^2}$, we shall adopt this restriction. In Fig. \ref{fig:shadow_I}, we compare the shadow radius's dependence on the magnetic charge $q$, for two distinct scenarios. Firstly, we consider $\gamma=\Lambda=0$ in order to evaluate how the shadow radius is affected solely by the magnetic charge. Secondly, we consider a scenario where $\gamma=10^{-11} M^{-1}$ and $\Lambda=10^{-20} M^{-2}$. At such values, these parameters exceed their corresponding (and independent) predicted values (i.e. in geometrized units: $\gamma \sim 10^{-17} M^{-1}$ \cite{Harada:2022edl} and $\Lambda \sim 10^{-33} M^{-2}$ \cite{Planck:2015fie, Planck:2018vyg}), and are therefore expected to affect the shadow radius significantly at the considered observer's distance. In both instances, we set the dimensionless parameter to $a=1$.

Inspecting Fig. \ref{fig:shadow_I} reveals that as the magnetic charge increases, the shadow radius of model I grows accordingly. In the case of $\gamma=\Lambda=0$, we find that $r_{sh}$ remains within the $2\sigma$ bounds at $q < 0.24M$. Conversely, when $\gamma=10^{-11} M^{-1}$ and $\Lambda=10^{-20} M^{-2}$, the $r_{sh}$ curve is shifted downwards, allowing higher charge values to be compatible with current observations. Notably, we find $r_{sh}$ remains within the $2\sigma$ bounds at $q < 0.54M$, resulting in less stringent constraints to the magnetic charge.

\begin{figure}[ht!]
    \includegraphics[width=\columnwidth]{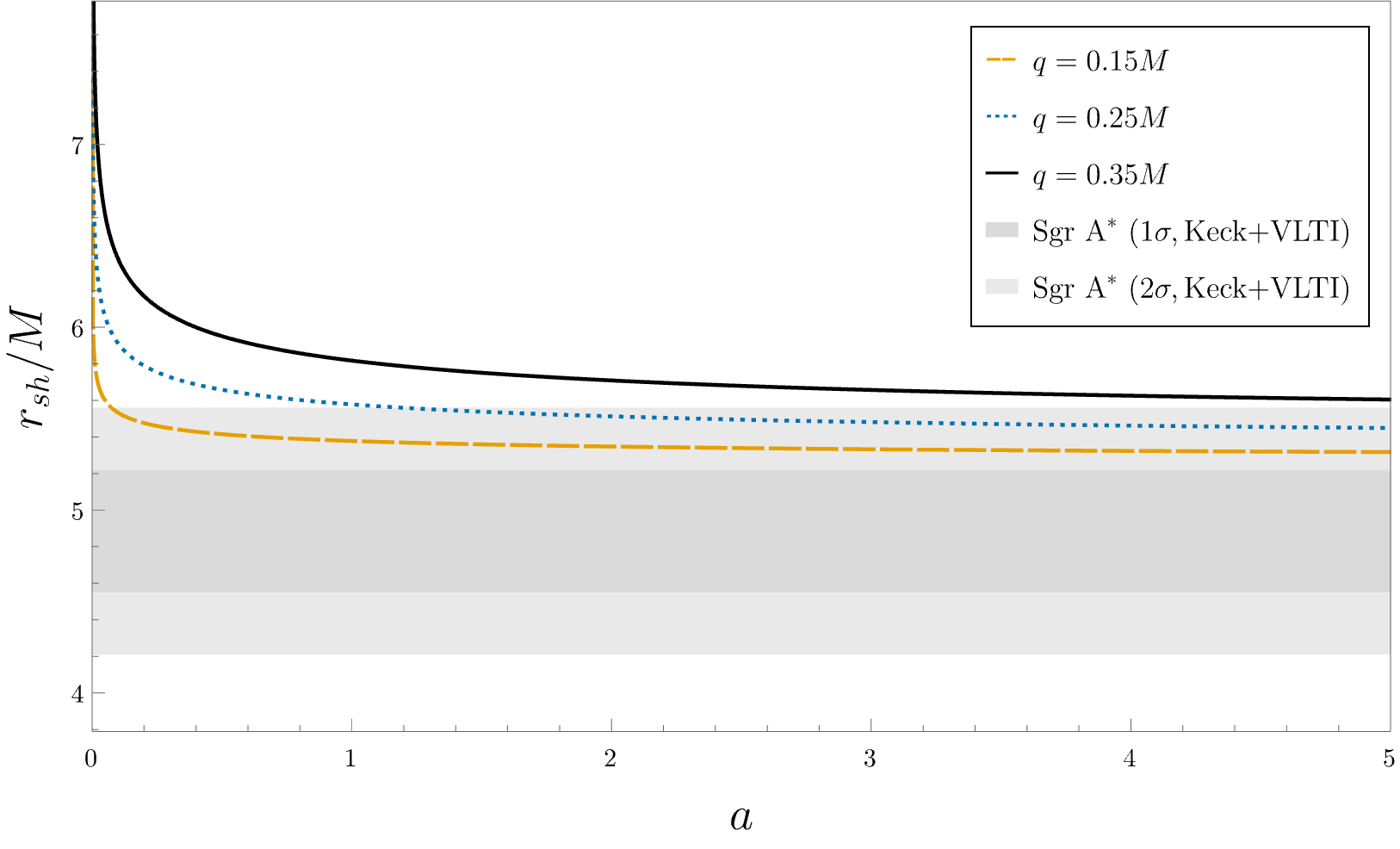}
    \caption{Shadow radius $r_{sh}$ of model I \eqref{eq:A(r)_I} as a function of the dimensionless parameter $a$ in three scenarios: $q=0.15M$ (dashed gold line); $q=0.25M$ (dotted blue line); $q=0.35M$ (black line). In all cases we consider $\gamma=\Lambda=0$. The shaded areas represent the confidence intervals of Sgr A*'s shadow radius at 1$\sigma$ (dark gray) and at 2$\sigma$ (light gray).}
    \label{fig:shadow_I_a}
\end{figure}

In Fig. \ref{fig:shadow_I_a}, we study the shadow radius's dependence on the dimensionless parameter $a$, for three values of the magnetic charge, always considering $\gamma=\Lambda=0$. Observing Fig. \ref{fig:shadow_I_a}, we note that the shadow radius decreases as $a$ increases. This behavior is more pronounced for smaller values of $a$, and becomes smoother as $q$ increases. Regardless, for $a \gg 1$ the shadow radius has a negligible dependence on this parameter. When $a \ll 1$ we find $r_{sh} \rightarrow +\infty$. Neglecting the influence of $\gamma$ and $\Lambda$, the $a$-parameter space with which $r_{sh}$ remains within the $2\sigma$ bounds is strongly influenced by the magnetic charge value, being confined to $q\gtrsim0.35M$. When $q\rightarrow 0$, $r_{sh}$ asymptotes to the Schwarzschild black hole shadow radius $3\sqrt{3}M$.

\subsection{Model II}\label{ssec:II_shadow}

\begin{figure}[ht!]
    \includegraphics[width=\columnwidth]{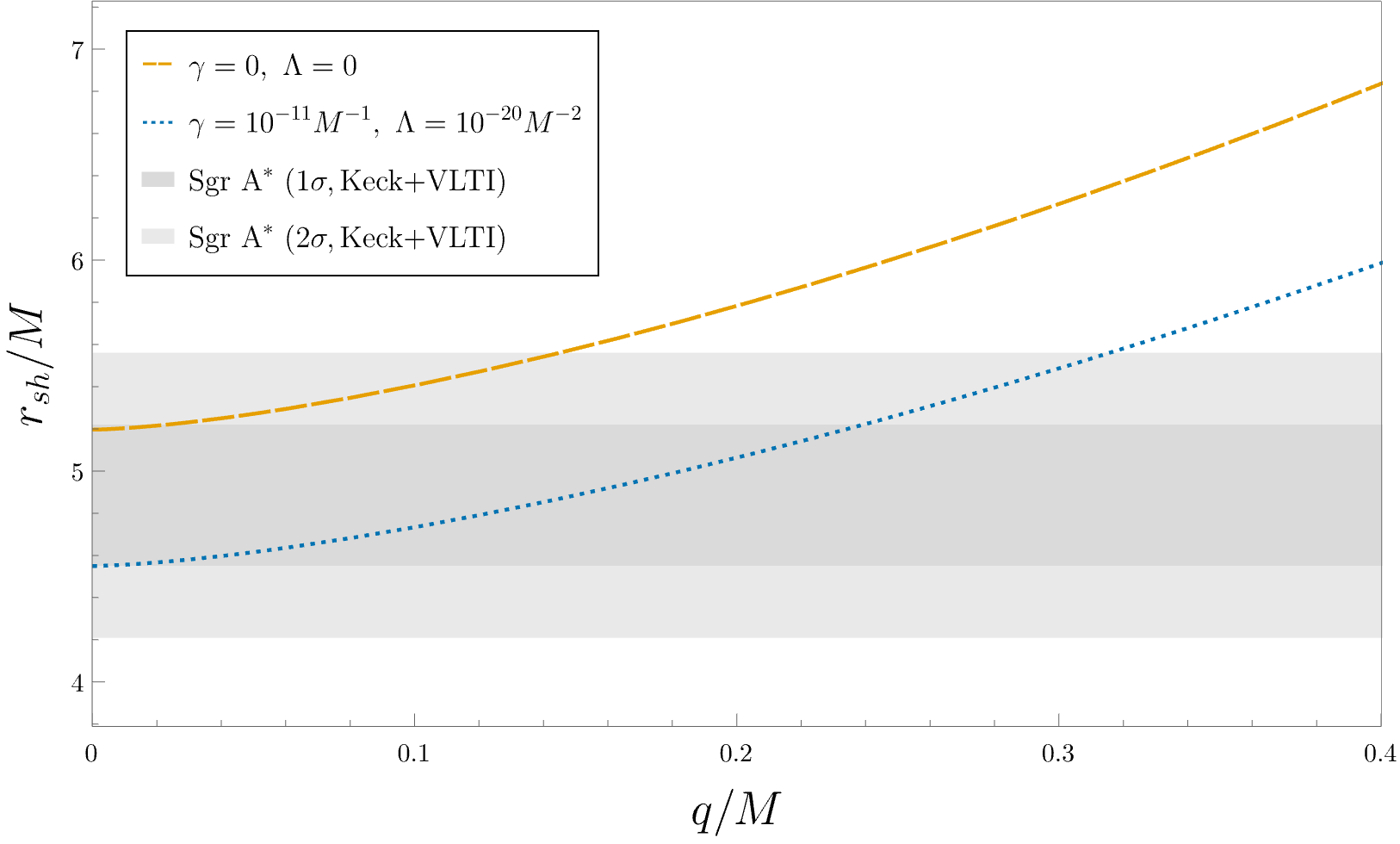}
    \caption{Shadow radius $r_{sh}$ of model II \eqref{eq:A(r)_II} as a function of the magnetic charge $q$ in two scenarios: $\gamma=\Lambda=0$ (dashed gold line); $\gamma=10^{-11} M^{-1}$ and $\Lambda=10^{-20} M^{-2}$ (dotted blue line). The shaded areas represent the confidence intervals of Sgr A*'s shadow radius at 1$\sigma$ (dark gray) and at 2$\sigma$ (light gray).}
    \label{fig:shadow_II}
\end{figure}

To compute the Model II's shadow radius, we repeat the previous procedure and obtain the effective geometry from Eqs. \eqref{eq:A(r)_II}, $C(r)=r^2$, and \eqref{eq:L_II}. The emergence of the $f_1$ parameter in the effective geometry leads us to impose the condition $f_1=\frac{\gamma}{2 q^2}$, as previously established. In Fig. \ref{fig:shadow_II}, we compare the shadow radius's dependence on the magnetic charge $q$, for two scenarios: $\gamma=\Lambda=0$; $\gamma=10^{-11} M^{-1}$ and $\Lambda=10^{-20} M^{-2}$. 

As can be seen in Fig. \ref{fig:shadow_II}, the shadow radius increases as the magnetic charge rises, for both considered cases. The effect is similar to that observed for model I (i.e. compare with Fig. \ref{fig:shadow_I}), albeit stronger in this model. Indeed, we find that when $\gamma=\Lambda=0$, $r_{sh}$ remains compatible with the $2\sigma$ bounds at $q < 0.14M$, which indicates that the shadow radius is more sensitive to the magnetic charge. When $\gamma=10^{-11} M^{-1}$ and $\Lambda=10^{-20} M^{-2}$, the $r_{sh}$ curve is vertically shifted and $r_{sh}$ remains within the $2\sigma$ bounds for $q < 0.32M$.

\subsection{Model III}\label{ssec:III_shadow}

\begin{figure}[ht!]
    \includegraphics[width=\columnwidth]{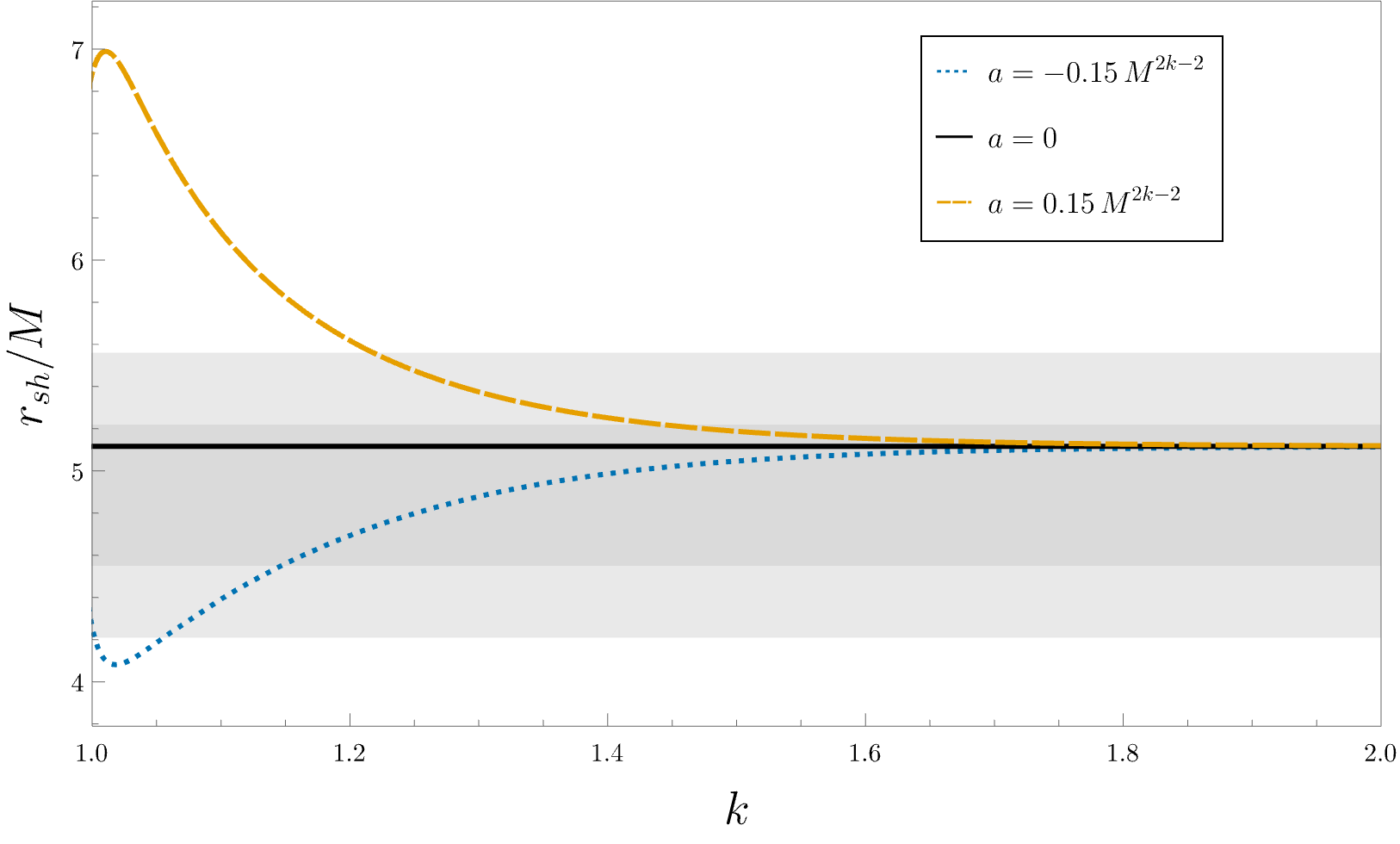}
    \caption{Shadow radius $r_{sh}$ of model III \eqref{eq:A(r)_III} as a function of the exponent $k$ in three scenarios: $a=0.15 M^{2k-2}$ (dashed gold line); $a=0$ (black line); $a=-0.15 M^{2k-2}$ (dotted blue line). The shaded areas represent the confidence intervals of Sgr A*'s shadow radius at 1$\sigma$ (dark gray) and at 2$\sigma$ (light gray). Fixed parameters: $q=0.3M$, $\gamma=10^{-17}$, $\Lambda=10^{-33}$.}
    \label{fig:shadow_III_k}
\end{figure}

Lastly, we obtain this model's effective geometry from Eqs. \eqref{eq:A(r)_III}, $C(r)=r^2$, and \eqref{eq:L_III}. This leaves us with a function characterized by six free parameters (i.e. $q, \gamma, \Lambda, k, a$ and $f_1$), further undermining the task of studying the shadow radius. To simplify our analysis, we focus our attention on this model's unique parameters, $a$ and the exponent $k$, and consider fixed values for $q$, $\gamma$, and $\Lambda$. We impose the restrictions $f_1=\frac{\gamma}{2 q^2}$ and $k \geq 1$, to ensure ${\cal L}(F)$ is linear in the weak field limit. In Fig. \ref{fig:shadow_III_k}, the shadow radius's dependence on the exponent $k$ is studied in three contexts: $a<0$, $a=0$, and $a>0$. We consider $q=0.3M$, and fix the cosmological constant and the Cotton parameter at $\gamma = 10^{-17} M^{-1}$, and $\Lambda = 10^{-33} M^{-2}$. Note that the choice of $a=0$ acts as a benchmark for the other two cases, since this effectively represents the Reissner-Nordstr\"{o}m black hole equivalent in Cotton gravity.
Observing Fig. \ref{fig:shadow_III_k}, we find that the shadow radius at $a=0$ is independent of the value of the exponent $k$, as expected. Its value coincides with that of a Reissner-Nordstr\"{o}m black hole with $q=0.3M$. In the case of $a>0$, the shadow radius slightly increases to a peak at $k \simeq 1.01$ before decreasing and converging to that of $a=0$, as the exponent $k$ increases in value; remaining compatible with the $2\sigma$ bounds at $k>1.22$. In contrast, when $a<0$ the shadow radius suffers a slight decrease to a minimum at $k \simeq 1.02$ before increasing and converging to that of $a=0$, as the exponent $k$ increases in value; remaining compatible with the $2\sigma$ bounds at $k>1.05$. To better understand this result, let us revisit Eq. \eqref{eq:A(r)_III}. In it, we find that $k$ is not only present as in the power of $r$ but also in an exponent of the magnetic charge (i.e. $q^{2k}$). Since the effective geometry retains this functional form, the contribution of the exponent $k$ to the shadow radius depends both on the magnetic charge and the parameter $a$. When $k=1$, the parameter $a$ plays the role of an effective charge. Thus, depending on whether it has a positive or negative value, its contribution exacerbates or hinders that of the magnetic charge, respectively. As the value of $k$ increases, the contributions of the magnetic charge and the parameter $a$ are less significant at large scales, causing the shadow radius to converge to the $a=0$ case. 

\begin{figure}[ht!]
    \includegraphics[width=\columnwidth]{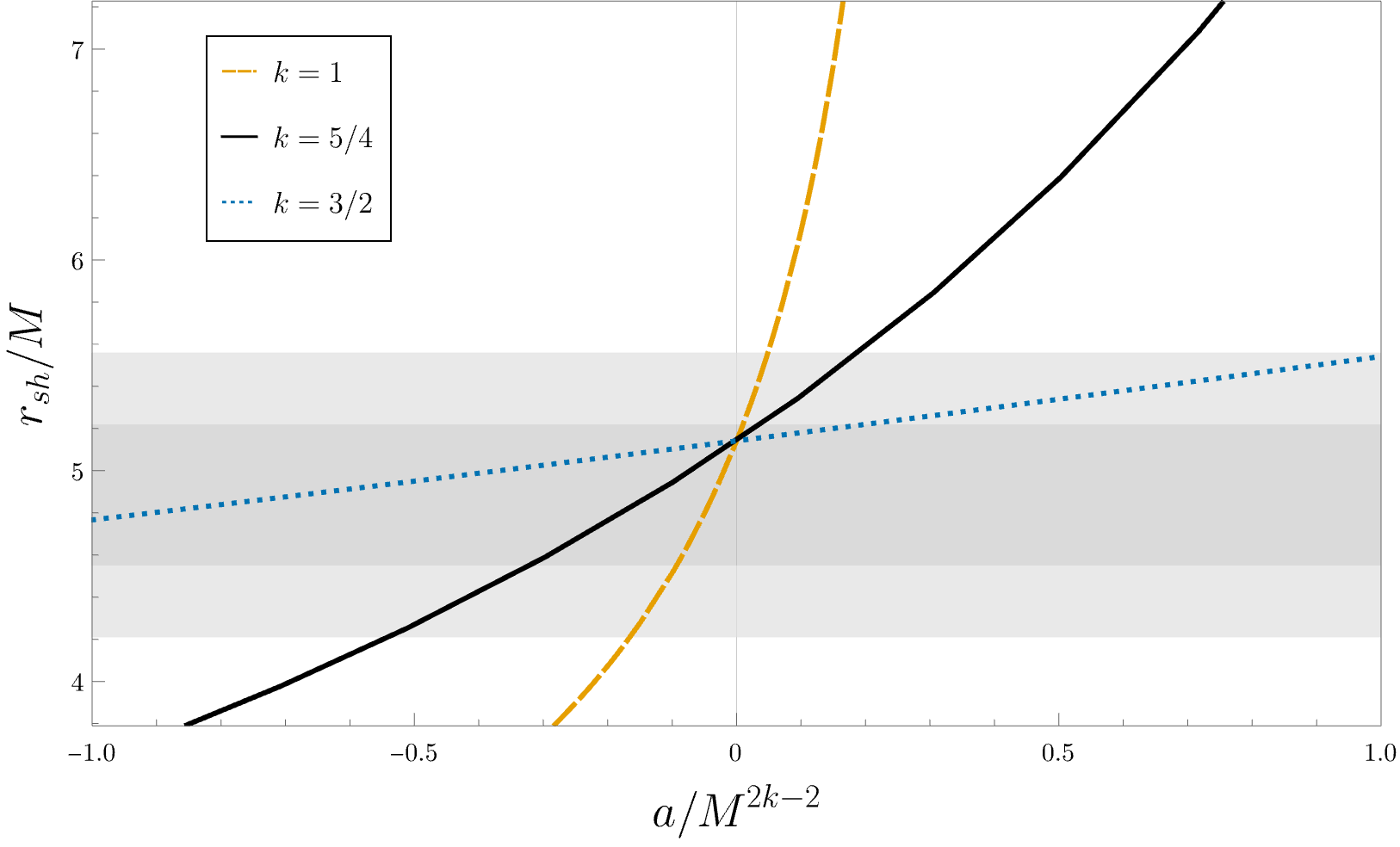}
    \caption{Shadow radius $r_{sh}$ of model III \eqref{eq:A(r)_III} as a function of the parameter $a$ in three scenarios: $k=1$ (dashed gold line); $k=\frac{5}{4}$ (black line); $a=\frac{3}{2}$ (dotted blue line). The shaded areas represent the confidence intervals of Sgr A*'s shadow radius at 1$\sigma$ (dark gray) and at 2$\sigma$ (light gray). Fixed parameters: $q=0.3M$, $\gamma=10^{-17}$, $\Lambda=10^{-33}$.}
    \label{fig:shadow_III_a}
\end{figure}

Next, we redirect our focus to the shadow radius' dependence on the parameter $a$. In Fig. \ref{fig:shadow_III_a}, we display $r_{sh}$ as a function of $a$, for three values of the exponent $k$: $k=1$, $k=\frac{5}{4}$, and $k=\frac{3}{2}$. In contrast with the previous case, the magnetic charge is independent of $a$. This allows us to examine how the shadow is affected solely by $a$ and $k$. We consider $q=0.3M$, $\gamma =10^{-17} M^{-1}$, and $\Lambda = 10^{-33} M^{-2}$.
In this context, we find that the compatibility of the parameter space of $a$ is strongly moderated by the exponent $k$. Based on Fig. \ref{fig:shadow_III_a}, the shadow radius increases in tandem with values of $a$ in all considered cases. This effect is more pronounced when $k=1$, where $r_{sh}$ is compatible with the $2\sigma$ bound between $-0.17  < a < 0.05 ;$, and less pronounced when $k=\frac{3}{2}$, where $r_{sh}$ remains fully consistent with the $2\sigma$ within the $-1.0 M < a < 1.0 M$ interval. 

\section{Summary and Conclusion}\label{sec:conc}

This paper investigated static spherically symmetric solutions within Cotton Gravity (CG), an extension of General Relativity (GR), coupled to nonlinear electrodynamics (NLED). We derived the CG field equations for a generic static and spherically symmetric metric with a purely magnetic NLED source, obtaining general expressions for ${\cal L}(r)$ and ${\cal L}_F(r)$ in terms of the metric function $A(r)$ and two integration constants. By solving these equations, we constructed magnetically charged geometries assuming arbitrary nonlinear Lagrangian densities. Reversing this approach, we determined the corresponding Lagrangian densities for specific metric functions and proposed three new spacetimes generalizing the neutral CG vacuum solution.

The first model, defined by ${\cal L}(F)=\frac{1}{1+aF}$, corresponds to the metric function \eqref{eq:A(r)_I}. The second model, with ${\cal L}(F)=F\left(\frac{1-F}{1+F}\right)$, is given by Eq. \;\eqref{eq:A(r)_II}. The third model, described by ${\cal L}(F)=F + aF^k$, corresponds to Eq.\;\eqref{eq:A(r)_III}, with a particular case at $k=\frac{3}{4}$ in \eqref{eq:A(r)_III_3/4}. All models extend the neutral CG vacuum solution \eqref{eq:CGvacBH} and are parameterized by $M$, $q$, $\gamma$, and $\Lambda$, with Model I depending additionally on a dimensionless parameter $a$, and Model III also depending on two parameters, $a$ and $k$.
The numerical analysis reveals a rich causal structure, including multi-horizon and horizonless configurations. Model I exhibits cases from naked singularities to infinite-horizon geometries. Model II supports up to two horizons: an event and a cosmological horizon. Model III, governed by $k$ and $a$, allows for configurations ranging from naked singularities to solutions with four horizons (two Cauchy, one event, and one cosmological horizon). All models satisfy the Maxwell weak field limit, provided $f_0=-\frac{\Lambda}{2}$ and $f_1=\frac{\gamma}{2 q^2}$ for models I and II. For model III, the additional constraints $k \geq 1$ or $f_1=\frac{\gamma-a(q^2/2)^{1/4}}{2 q^2}$ with $k=1/4$ are required. Additionally, all models were tested against the classical (null, weak, strong, and dominant) energy conditions. The radial coordinates at which these conditions are satisfied were identified for all models. In this regard, it was found that all models violate at least one of the energy conditions, with such violations taking place inside or outside the event horizon depending on the values of the free parameters defining a given geometry. These results signal that, for certain parameter values—specifically those for which the energy conditions are respected outside the event horizon—the models appear to be classically viable and well-behaved for an external observer, being devoid of any troublesome pathologies such as instabilities and/or causality violations outside the event horizon.

In Section \ref{sec:BHshadows}, we analyzed the parameter-dependent shadow radius of the three models, incorporating the effective geometry governing photon propagation due to NLED. For a static observer, we accounted for the non-asymptotically flat nature of these spacetimes and compared theoretical predictions with EHT observations, using the $1\sigma$ and $2\sigma$ bounds of Sgr A*'s shadow to constrain the parameter space \cite{Vagnozzi:2022moj}.  
Our results highlight distinct parameter space constraints among the models. For models I and II, the shadow radius increases with the magnetic charge $q$, with a stronger effect in model II. In model III, the shadow radius scales directly with $a$, while $k$ modulates this dependence, causing it to converge to the $a=0$ case as $k$ increases. Overall, these models remain compatible with current observational constraints, supporting their potential as descriptions of Sgr A*'s shadow.  

In summary, coupling NLED with CG has led to three new charged, spherically symmetric black hole solutions, extending classical geometries while maintaining the core features of this alternative gravity theory. While the analysis of the energy conditions is an important first step towards assessing the viability of these models, it is not sufficient on its own, and several important consistency checks remain as possible avenues of future research. For example, the study of the unitarity and causality conditions \cite{Shabad:2011hf} could yield more profound insights into the presence of tachyons and ghost-like states in these models. This is particularly relevant in the context of NLED frameworks coupled to GR (and thereby extensions of GR), as static, spherically symmetric magnetic solutions of the GR-NLED field equations have been shown to always violate the causality and unitarity conditions at some point near a regular center \cite{Bronnikov:2022ofk}.

Additionally, studying the metric stability of these solutions under linear perturbations or via the study of their quasinormal modes and analyzing their thermodynamic properties (e.g., entropy, specific heat, temperature) would provide a better understanding of these solutions' stability and whether they are physically realizable. In this regard, it has been recently noted that gravitational models based on a tensor with rank greater than 2 (i.e., such as Cotton gravity) do not allow for an interpretation of parameters such as the black hole mass in terms of conserved quantities \cite{Altas:2025pvg}. Additionally, although this limitation can be circumvented when studying the black hole shadow by resorting to measurements of the local mass (see the discussion in subsection C.T. in \cite{Vagnozzi:2022moj} and references therein), this remains an obstacle with regards to analyzing these models' thermodynamic properties and spacetime stability, which rely on a well-defined notion of mass.

Future work could explore their regularization, dyonic extensions, black bounce configurations, and additional models based on well-motivated NLED frameworks (e.g., \cite{Born:1934gh,Born:1934ji,Heisenberg:1936nmg,Bandos:2020jsw}). 
Furthermore, NLED offers a refined framework that may alleviate some of CG's intrinsic issues, opening avenues for stability analysis and thermodynamic investigations. While the shadow radius constraints from EHT provide a preliminary assessment, further observational tests will be essential to impose tighter parameter bounds and distinguish these models from other viable alternatives.


\acknowledgments{
FSNL acknowledges support from the Funda\c{c}\~{a}o para a Ci\^{e}ncia e a Tecnologia (FCT) Scientific Employment Stimulus contract with reference CEECINST/00032/2018, and funding through the research grants UIDB/04434/2020, UIDP/04434/2020 and PTDC/FIS-AST/0054/2021.
MER thanks CNPq for partial financial support.  This study was supported in part by the Coordenção de Aperfeioamento de Pessoal de Nível Superior - Brazil (CAPES) - Financial Code 001.  }


%

\end{document}